%% file: ms.tex
\documentclass[twocolumn]{aastex63}

\newcounter{daggerfootnote}
\newcommand*{\daggerfootnote}[1]{%
    \setcounter{daggerfootnote}{\value{footnote}}%
    \renewcommand*{\thefootnote}{\fnsymbol{footnote}}%
    \footnote[2]{#1}%
    \setcounter{footnote}{\value{daggerfootnote}}%
    \renewcommand*{\thefootnote}{\arabic{footnote}}%
    }

\usepackage{amsmath}
\usepackage{color}

\newcommand{\mpchance}{P_{\rm chance}}
\newcommand{\pchance}{$\mpchance$}
\newcommand{\achance}{0.05}

\newcommand{\pypeit}{{\tt PypeIt}}

\newcommand{\mkms}{{\rm km \, s^{-1}}}

\newcommand{\mmstar}{M_\star}
\newcommand{\mstar}{$\mmstar$}
\newcommand{\mebv}{E(B-V)}
\newcommand{\ebv}{$\mebv$}

\newcommand{\frbsix}{21h22m58.91s, $-$79d23m51.3s} 
\newcommand{\frbsev}{21h57m40.68s, $-$80d21m28.8s} 
\newcommand{\frbsef}{12h15m55.12s, $-$13d01m15.7s} 
\newcommand{\frbten}{21h33m24.373s, $-$54d44m51.43s} 
\newcommand{\frbtut}{15h18m49.54s, $+$12d22m36.8s} 

\newcommand{\bcandse}{21h22m58.28s,$-$79d23m50.1s} 
\newcommand{\fcandse}{21h22m58.97s,$-$79d23m51.7s} 
\newcommand{\acandst}{J042017.71$+$734222.9} 
\newcommand{\bcandst}{J042017.87$+$734224.4} 
\newcommand{\hgsvev}{21h57m40.60s,$-$80d21m29.25s} 
\newcommand{\hgsvf}{J121555.0941$-$130116.004} 
\newcommand{\hgoo}{J213324.44$-$544454.65} 
\newcommand{\hgBoo}{J213323.65$-$544453.6} 
\newcommand{\sone}{J134815.44$+$722814.72} 
\newcommand{\stwo}{J134815.74$+$722805.9} 
\newcommand{\hgtut}{J151849.52$+$122235.82} 

\newcommand{\ntot}{13} 
\newcommand{\nsampA}{10} 

\newcommand{\mzspec}{z_{\rm spec}}
\newcommand{\zspec}{$\mzspec$}
\newcommand{\mzphot}{z_{\rm phot}}
\newcommand{\zphot}{$\mzphot$}

\newcommand{\mReff}{R_{\rm eff}}
\newcommand{\Reff}{$\mReff$}

\newcommand{\mdmunits}{{\rm pc\,cm^{-3}}} 
\newcommand{\dmunits}{$\mdmunits$}
\newcommand{\mdmcosmic}{{\rm DM}_{\rm cosmic}}
\newcommand{\dmcosmic}{$\mdmcosmic$}
\newcommand{\mdmacosmic}{\langle {\rm DM}_{\rm cosmic} \rangle}

\newcommand{\mdmfrb}{{\rm DM}_{\rm FRB}}

\newcommand{\mdmhost}{{\rm DM}_{\rm host}}
\newcommand{\dmhost}{$\mdmhost$}
\newcommand{\mdmmwhalo}{{\rm DM}_{\rm MW,halo}}
\newcommand{\dmmwhalo}{$\mdmmwhalo$}
\newcommand{\mdmmwism}{{\rm DM}_{\rm MW,ISM}}

\newcommand{\oiii}{O\,{\sc iii}}
\newcommand{\oii}{O\,{\sc ii}}
\newcommand{\nii}{N\,{\sc ii}}
\newcommand{\sii}{S\,{\sc ii}}
\newcommand{\ha}{H$\alpha$}
\newcommand{\hb}{H$\beta$}

\newcommand{\mpks}{P_{\rm KS}}
\newcommand{\pks}{$\mpks$}

\begin{document}

\title{Host Galaxy Properties and Offset Distributions of Fast Radio Bursts: Implications for their Progenitors}

\correspondingauthor{Kasper E. Heintz}
\email{keh14@hi.is}

\author[0000-0002-9389-7413]{Kasper E. Heintz}
\affil{Centre for Astrophysics and Cosmology, Science Institute, University of Iceland, Dunhagi 5, 107 Reykjav\'ik, Iceland}

\author{J. Xavier Prochaska}
\affil{University of California - Santa Cruz, 1156 High St., Santa Cruz, CA, USA 95064}
\affil{Kavli Institute for the Physics and Mathematics of the Universe (Kavli IPMU), 5-1-5 Kashiwanoha, Kashiwa, 277-8583, Japan}

\author[0000-0003-3801-1496]{Sunil Simha}
\affil{University of California - Santa Cruz, 1156 High St., Santa Cruz, CA, USA 95064}

\author{Emma Platts}
\affil{High Energy Physics, Cosmology \& Astrophysics Theory (HEPCAT) group, Department of Mathematics and Applied Mathematics, University of Cape Town, South Africa}

\author[0000-0002-7374-935X]{Wen-fai Fong}
\affil{Center for Interdisciplinary Exploration and Research in Astrophysics and Department of Physics and Astronomy, Northwestern University, 2145 Sheridan Road, Evanston, IL 60208-3112, USA}

\author{Nicolas Tejos}
\affil{Instituto de F\'isica, Pontificia Universidad Cat\'olica de Valpara\'iso, Casilla 4059, Valpara\'iso, Chile}

\author[0000-0003-4501-8100]{Stuart D. Ryder}
\affil{Department of Physics \& Astronomy, Macquarie University, NSW 2109, Australia}
\affil{Astronomy, Astrophysics and Astrophotonics Research Centre, Macquarie University, Sydney, NSW 2109, Australia}

\author[0000-0002-2059-0525]{Kshitij Aggarwal}
\affil{Department of Physics and Astronomy, West Virginia University, Morgantown, WV 26506, USA}
\affil{Center for Gravitational Waves and Cosmology, West Virginia University, Chestnut Ridge Research Building, Morgantown, WV, USA}

\author[0000-0003-3460-506X]{Shivani Bhandari}
\affil{Australia Telescope National Facility, CSIRO Astronomy and Space Science, PO Box 76, Epping, NSW 1710, Australia}

\author[0000-0002-8101-3027]{Cherie K. Day}
\affil{Centre for Astrophysics and Supercomputing, Swinburne University of Technology, Hawthorn, VIC 3122, Australia}
\affil{Australia Telescope National Facility, CSIRO Astronomy and Space Science, PO Box 76, Epping, NSW 1710, Australia}

\author{Adam T. Deller}
\affil{Centre for Astrophysics and Supercomputing, Swinburne University of Technology, Hawthorn, VIC 3122, Australia}

\author{Charles D. Kilpatrick}
\affil{Department of Astronomy and Astrophysics, University of California, Santa Cruz, CA 95064, USA}

\author{Casey J. Law}
\affil{Cahill Center for Astronomy and Astrophysics, MC 249-17 California Institute of Technology, Pasadena, CA 91125, USA}

\author{Jean-Pierre Macquart$^\dagger$}
\affil{International Centre for Radio Astronomy Research, Curtin University, Bentley WA 6102, Australia}
\daggerfootnote{Deceased}

\author{Alexandra Mannings}
\affil{University of California - Santa Cruz, 1156 High St., Santa Cruz, CA, USA 95064}

\author[0000-0003-1483-0147]{Lachlan J. Marnoch}
\affil{Department of Physics \& Astronomy, Macquarie University, NSW 2109, Australia}
\affil{Astronomy, Astrophysics and Astrophotonics Research Centre, Macquarie University, Sydney, NSW 2109, Australia}
\affil{Australia Telescope National Facility, CSIRO Astronomy and Space Science, PO Box 76, Epping, NSW 1710, Australia}

\author{Elaine M. Sadler}
\affil{Australia Telescope National Facility, CSIRO Astronomy and Space Science, PO Box 76, Epping, NSW 1710, Australia}
\affil{Sydney Institute for Astronomy, School of Physics A28, The University of Sydney, NSW 2006, Australia}

\author{Ryan M. Shannon}
\affil{Centre for Astrophysics and Supercomputing, Swinburne University of Technology, Hawthorn, VIC 3122, Australia}

\shorttitle{FRB host galaxies}
\shortauthors{K. E. Heintz et al.}

\received{\today}
\revised{\today}
\accepted{\today}
\submitjournal{ApJ}

\begin{abstract}
    We present observations and detailed characterizations of five new host galaxies of fast radio bursts (FRBs) discovered with the Australian Square Kilometre Array Pathfinder (ASKAP) and localized to $\lesssim 1''$. 
    Combining these galaxies with FRB hosts from the literature, we 
    introduce criteria based on the probability of chance coincidence
    to define a subsample of 10 highly confident associations (at $z=0.03-0.52$), 3 of which correspond to known repeating FRBs.
    Overall, the FRB-host galaxies exhibit a broad, continuous range of color ($M_u-M_r = 0.9 - 2.0$), stellar mass ($M_\star = 10^{8} - 6\times 10^{10}\,M_{\odot}$), and star-formation rate (${\rm SFR} = 0.05 - 10\,M_{\odot}\,{\rm yr}^{-1}$) spanning the full parameter space occupied by $z<0.5$ galaxies. 
    However, they do not track the color-magnitude, SFR$-M_\star$, nor BPT diagrams of field galaxies 
    surveyed at similar redshifts. There is an excess of ``green valley'' galaxies and an excess of emission-line ratios indicative of a harder radiation field than that generated
    by star-formation alone.
    From the observed stellar mass distribution, we 
    rule out the hypothesis that FRBs strictly track stellar
    mass in galaxies ($>99\%$\,c.l.).
    We measure a median offset of 3.3\,kpc from the FRB to the estimated center of the host galaxies and
    compare the host-burst offset distribution and other properties with the distributions of long- and short-duration gamma-ray bursts (LGRBs and SGRBs), core-collapse supernovae (CC-SNe), and SNe Ia.
    This analysis rules out galaxies hosting LGRBs (faint, star-forming galaxies) as common hosts for FRBs ($>95$\%\,c.l.). Other transient channels (SGRBs, CC-, and SNe Ia) have host-galaxy properties and offsets consistent with the FRB distributions. All of the data and derived quantities are made publicly available on a dedicated website and repository.
\end{abstract}


\section{Introduction} \label{sec:intro}

The transients classified as fast radio bursts (FRBs) and their progenitors constitute one of the major puzzles in contemporary astrophysics \citep[see][for recent reviews]{Cordes19,Petroff19}. FRBs are brief ($\sim 1$\,ms), but bright ($> 1$\,Jy\,ms) radio-pulse events, similar in nature to pulsars, although their extragalactic origin \citep{Thornton13} implies much higher energies. Despite being first detected more than a decade ago \citep{Lorimer07}, the physical engines powering FRBs still remain a mystery, but a plethora of origins has been proposed \citep[see e.g.][for a compendium]{Platts19}.

Nevertheless, FRBs have already been demonstrated to be powerful cosmological probes. Similar to how UV or optically bright cosmic beacons such as quasars and gamma-ray burst (GRB) afterglows have been paramount in the study of the interstellar and intergalactic gas properties at high redshifts \citep{Wolfe05,Fynbo09}, FRBs have revolutionized the studies of the ``cosmic web'' between galaxies \citep{Macquart20,Simha20}, the diffuse ionized gas in extragalactic halos \citep{McQuinn14,Prochaska19a,Prochaska19b}, and the interstellar and circumgalactic media of their hosts \citep{Tendulkar17,Chittidi20}. Most notably, FRBs can be used to provide a census of the baryonic content that is in a highly diffuse state and therefore difficult to detect with any other approach \citep{Macquart20}.

Until recently, the main issue hindering any significant progress has been the generally poor localizations of the events. The first decade of FRB searches was undertaken with telescopes that had localization regions  $\gg 1$\,arcmin$^2$. This is inhibited by the seeming lack of ``afterglows'' analogous to those observed for GRBs \citep{Petroff17,Bhandari18,Chen20} and associated supernova-like transient counterparts \citep{Marnoch20}. A precise localization ($\sim 1''$) of the burst itself is thus required to robustly identify the associated host galaxy \citep{Eftekhari17}.

The first unique identification of an FRB-host galaxy was based on direct interferometric localization of the repeat bursts from FRB\,121102 \citep{Spitler16}. Follow-up observations revealed a faint, actively star-forming (SF), low-mass galaxy at $z=0.1927$ \citep{Chatterjee17,Tendulkar17}. The resemblance to the hosts of long-duration GRBs and superluminous supernovae (SLSNe) promoted ``young'' flaring magnetar models as the origin of the repeat bursts \citep[e.g.][]{Metzger17,Margalit18}. However, it is now clear that the host galaxy of FRB\,121102 is anomalous compared to other FRB hosts \citep[e.g.][]{Bannister19,Li19,Bhandari20a}. Recently, another repeating FRB, FRB\,180916, was localized to an SF region in a nearby spiral galaxy \citep{Marcote20}, showing properties in stark contrast to the host of FRB\,121102.  

The Commensal Real-Time ASKAP Fast Transients \citep[CRAFT;][]{Macquart10} survey has operated the Australian Square Kilometre Array Pathfinder (ASKAP) in incoherent-sum (ICS) mode since 2018, and now routinely provides $\sim$arcsecond localizations of single-pulse FRBs. This led to the discovery of the first two host galaxies associated with apparently one-off FRBs \citep{Bannister19,Prochaska19b}, and based on the first preliminary study of ASKAP-detected FRBs \citep[][see also \citealt{Li20}]{Bhandari20a}, it is now clear that the majority of FRB hosts are instead massive galaxies with older stellar populations. This suggests that FRBs reside in diverse environments, even for the proposed subpopulation of repeating bursts.
The progenitors of FRBs (and astronomical transients in general) are likely linked to specific stellar populations and galactic environments, so detailed characterizations of their host galaxies allow us to constrain the nature of these events and their likely progenitor channels \citep[akin to how the host properties of GRBs aided in constraining their progenitors, e.g.][]{Fruchter06,Yoon06}. 

In this paper, we present the first comprehensive and statistical analyses of the population of galaxies hosting FRBs. These include detailed characterizations of five new host galaxies of accurately localized FRBs detected by ASKAP. Combined with all previously identified FRB hosts reported in the literature, our sample comprises a total of 13 host galaxies. We measure the physical properties of the majority of the FRB hosts in our sample based on existing and newly obtained spectroscopic and photometric data.

Throughout the paper, we distinguish between host galaxies of repeating FRBs and apparently nonrepeating, one-off bursts to investigate any distinct characteristics between the host populations of the two apparent types of FRBs. We first compare the observed FRB-host properties to those of field galaxies to examine how the FRB hosts are drawn from the underlying galaxy population. We then investigate any connections between the FRB-host properties and host-burst offset distributions to those of other astronomical transients such as long-duration GRBs (LGRBs), short-duration GRBs (SGRBs), core-collapse supernovae (CC-SNe) and SNe Ia. Recently, \citet{Li20} and \citet{Bhandari20a} analyzed a sample of five and six FRB hosts, respectively, and found that their physical properties are most consistent with those of SGRBs and SNe Ia, excluding models in which the majority of FRBs originate from SLSNe/LGRB progenitors or active galactic nuclei (AGNs).
Here, we leverage our larger sample to further narrow down and provide stronger constraints on the most likely progenitor channels for the majority of FRBs.

We have structured the paper as follows: in Section~\ref{sec:sample} we define the FRB-host galaxy sample(s) and present the new host-galaxy observations of the ASKAP-localized FRBs characterized here. We detail the modeling of the host-galaxy properties in Section~\ref{sec:analysis} and compare the typical host-galaxy environments to field-selected galaxies in Section~\ref{sec:hostprop}. In Section~\ref{sec:frbprog} we compare the FRBs to other types of astronomical transients and discuss what the implications of our results are on the most likely FRB progenitor channels. We conclude and summarize our work in Section~\ref{sec:conc}. Throughout the paper, we assume the concordance cosmological model, with $\Omega_{\rm m} = 0.308$ and $H_0 = 67.8$\,km\,s$^{-1}$\,Mpc$^{-1}$ \citep{Planck16}.

\section{Sample and Observations}
\label{sec:sample}

In collaboration with the CRAFT \citep{Macquart10} and {\it realfast} \citep{Law18} surveys, we have as part of the Fast and Fortunate for FRB Follow-up (F$^4$)\footnote{\url{ucolick.org/f-4}} collaboration endeavored to obtain dedicated photometric and spectroscopic follow-up observations of all $\sim$arcsecond-localized FRBs. These provide a secure identification of the associated host galaxies and allow us to derive their main physical properties. All the observational data products are available on the FRB GitHub repository\footnote{\url{https://github.com/FRBs/FRB}}, in addition to a large suite of FRB-related scripts. As a front-end to these data repositories, we have also launched an online FRB-host galaxy database\footnote{\url{https://frbhosts.org}}, with the goal of collecting and sharing all currently known and future FRB hosts and their basic properties.

In this section, we describe the identification of FRB-host galaxies and define a set of sample criteria to describe the robustness 
of the host associations. We then present the new observations of five FRB-host galaxies and compile all previously known FRB hosts reported in the literature, all considered in our meta analysis. At the end, we summarize the overall sample properties.


\subsection{Host-galaxy Associations}
\label{ssec:associate}

An FRB signal alone cannot directly establish the redshift of the source, and one relies on an
association with a host galaxy for a precise measurement. 
To date and in this work,
the association of the FRB with a host galaxy is primarily based on probabilistic arguments given their position relative to coincident or nearby galaxies. Following standard practice for other transients \citep[e.g.][for GRBs]{Bloom02,Blanchard16}, one may estimate the probability of a chance coincidence ($P_{\rm chance}$) based on the angular offset, $\theta$, of the FRB position from the galaxy centroid,
the uncertainty of the FRB localization,
and the galaxy's apparent magnitude. 
Further work may adopt additional properties and
priors for establishing associations.

\input{tab_sample}

The derivation of $P_{\rm chance}$ is based on galaxy number counts and captures the fact that apparently faint galaxies are more common on the sky. We adopt the formalism developed by \cite{Bloom02}, derived from optical galaxy number counts \citep{Hogg97}, which gives the number density of galaxies brighter than apparent $r$-band magnitude $m_r$ (not taking into account clustering of galaxies), as

\begin{equation}
\begin{split}
    \Sigma(\leq m_r) = &\frac{1}{3600^2 \times 0.334\,\log_{e}(10)} \\ \times &10^{0.334 (m_{r} - 22.963) + 4.320}\,\,{\rm arcsec^{-2}} \;.
\end{split}
\end{equation}
We then calculate the probability of chance coincidence, given by

\begin{equation}
    \mpchance = 1 - \exp(-\eta) \;,
\end{equation}

\noindent where $\eta \equiv \pi \theta^2 \Sigma(\leq m_r)$. We report the estimated chance probabilities of each of the FRB-host galaxies in Table~\ref{tab:sample}. Here, we also provide the association radius $\delta x$, representing the offset from a given galaxy with $r$-band magnitude $m_r$ within which the FRB can be securely associated with the galaxy \citep{Tunnicliffe14}.

In previous works, we estimated the probability of chance
coincidence with an empirical approach \citep{Bannister19}
and reported $\mpchance < 10^{-3}$ for the first
well-localized ASKAP-detected FRBs \citep[e.g.][]{Bannister19,Prochaska19b}. The formalism described
above yields consistent results. We note that \citet{Eftekhari17} have developed a similar framework to quantify the robustness of the FRB-host galaxy associations with a more recent number count estimation. This generally provides lower chance probabilities;
here, we use the formalism described above to be 
more conservative. In this work, we also estimate the uncertainty on the offsets from the FRB to the host galaxy center by integrating over the FRB localization ellipse.

Our approach is designed to (i) minimize the 
deleterious effect of false positives on this somewhat small sample of
events and (ii) define a high-confidence sample that can be used
in future analyses to generate priors for a full Bayesian analysis.
To do this, we define four subsamples based solely on \pchance\ and the 
quality of the galaxy redshift estimation.
These are:

\begin{itemize}
    \item Sample A: The host-galaxy association is considered highly probable
    ($\mpchance < \achance$) based on the FRB localization and galaxy
    photometry. The galaxy has a spectroscopically confirmed redshift \zspec.
    \item Sample B: Same as Sample A, except that only a photometric
    redshift \zphot\ has been estimated.
    \item Sample C: The host-galaxy association is less secure due to a poor FRB localization, multiple host candidates, and/or because
    additional priors were adopted in the association
    \citep[e.g. the Macquart DM$-z$ relation;][]{Macquart20}.
    A spectroscopic redshift \zspec\ has been measured.
    \item Sample D: Same as Sample C, except that only a photometric
    redshift \zphot\ has been measured.
\end{itemize}

We consider all the FRB hosts compiled in this work throughout the paper but caution about the potential pitfalls of the uncertain host-galaxy identifications where relevant. For the statistical analyses we only consider the FRBs in Sample~A. In the following section we introduce all of the candidate FRB-host galaxies and enumerate the number in each sample type.

\subsection{FRB-host Galaxy Observations}
\label{ssec:new_frbs}

In continuation of the first four FRBs detected and accurately localized by ASKAP/CRAFT \citep[presented in][]{Bhandari20a}, we here report the observations and basic properties of five more recent FRB-host galaxies: those of FRBs\,190611, 190711, 190714, 191001, and 200430.

\subsubsection{FRB\,190611}
\label{ssec:190611}

On UT 2019 June 11 at 05:45:43.3, the ASKAP telescope recorded FRB\,190611 as reported by \citet{Macquart20}, who also briefly described its host-galaxy candidates. The FRB position is at R.A., decl. ($\alpha,\delta$) = \frbsix\ (J2000), with an uncertainty of $\sigma_{\alpha, \delta} = 0\farcs7, 0\farcs7$. 

We obtained deep Gemini-S/GMOS images in the 
$r$ and $i$ bands (the latter shown in Figure~\ref{fig:images}) revealing a bright source ($r = 22.65$\,mag) approximately $2\farcs0$ to the north-west at $\alpha,\delta = $ \bcandse\ (J2000), identified as the host galaxy by \citet{Macquart20}. We do not detect any significant structure (e.g., spiral arms) and measure an effective half-light radius of $R_{\rm eff} = 0\farcs40$. We also tentatively detect a considerably fainter source coincident within
the FRB error ellipse ($r \approx 26$\,mag; at \fcandse) at a smaller offset of $0\farcs43$ from the FRB position.  
We estimate chance probabilities for the two galaxies to be unrelated to the FRB host of $\mpchance = 0.017, 0.10$ for the bright and faint galaxy, respectively. Given the only tentative detection of the faint source and that the bright source has $\mpchance \approx 2\%$, we consider the more clearly offset, bright galaxy to be the host of FRB\,190611 and place it in our primary Sample~A.

Spectroscopy of this host-galaxy candidate with the FORS2 instrument on the ESO Very Large Telescope (VLT) was reduced
using the \pypeit\ reduction package \citep{pypeit},
which optimally extracts a 1D spectrum from the flat-fielded
and sky-subtracted 2D spectral image. We additionally performed a 2D coaddition of the spectra presented in \citet{Macquart20}. This
yields a spectroscopic redshift of $\mzspec = 0.3778$ based on the \ha, \hb, and [\oiii] line features. At this redshift, the physical projected offset of the FRB from the bright galaxy centroid is $\approx 11$\,kpc.


\subsubsection{FRB\,190711}
\label{ssec:190711}

On UT 2019 July 11 at 01:53:41.1, the ASKAP telescope recorded FRB\,190711 as reported by \citet{Macquart20},
who also provided a brief description of its host galaxy. The FRB position is at $\alpha,\delta$ = \frbsev\ (J2000), with an uncertainty of $\sigma_{\alpha, \delta} = 0\farcs4, 0\farcs3$ \citep{Day20}. This FRB has subsequently been found to repeat \citep{Kumar20}.

\begin{figure*}[!t]
\centering
    \includegraphics[width=5.7cm]{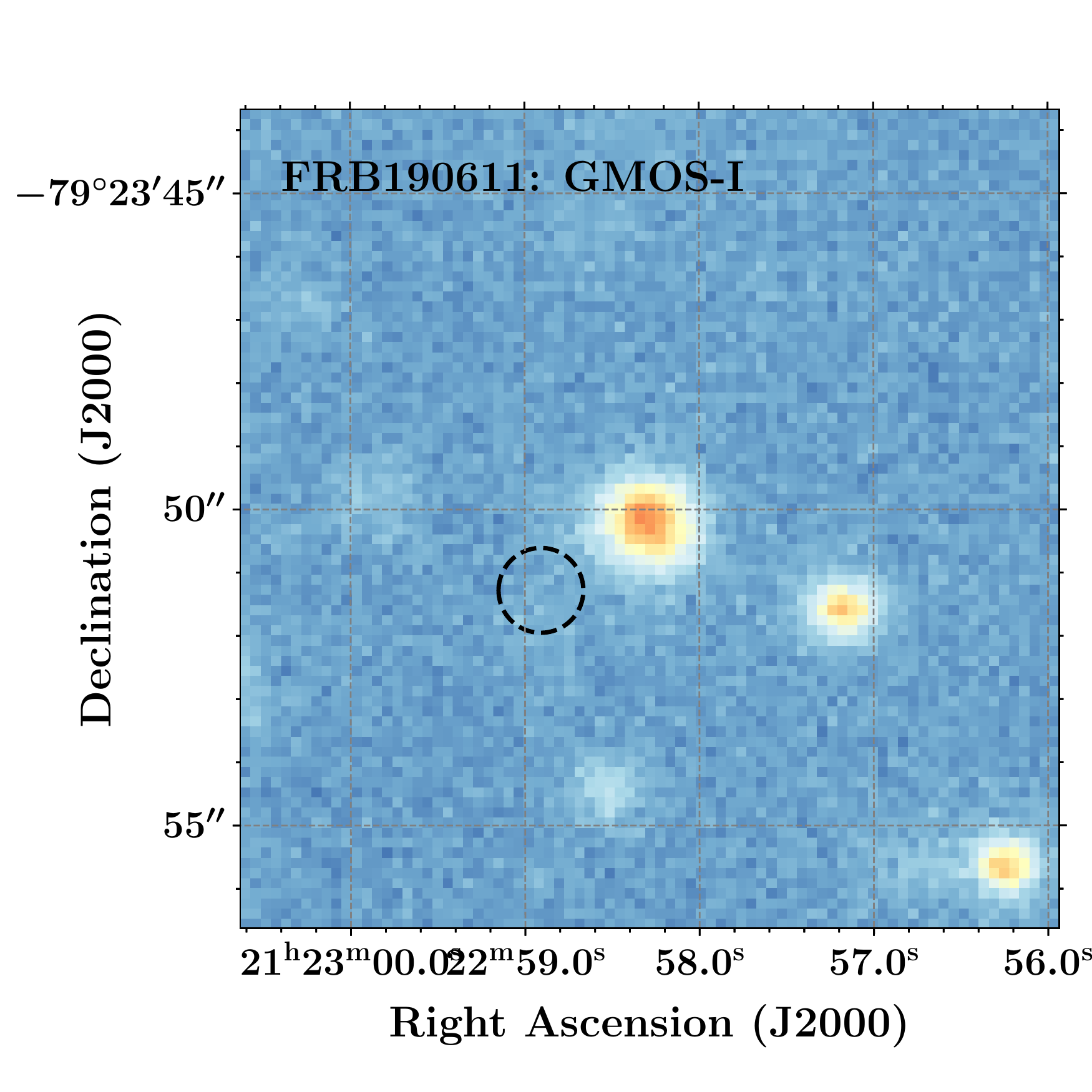}
    \includegraphics[width=5.7cm]{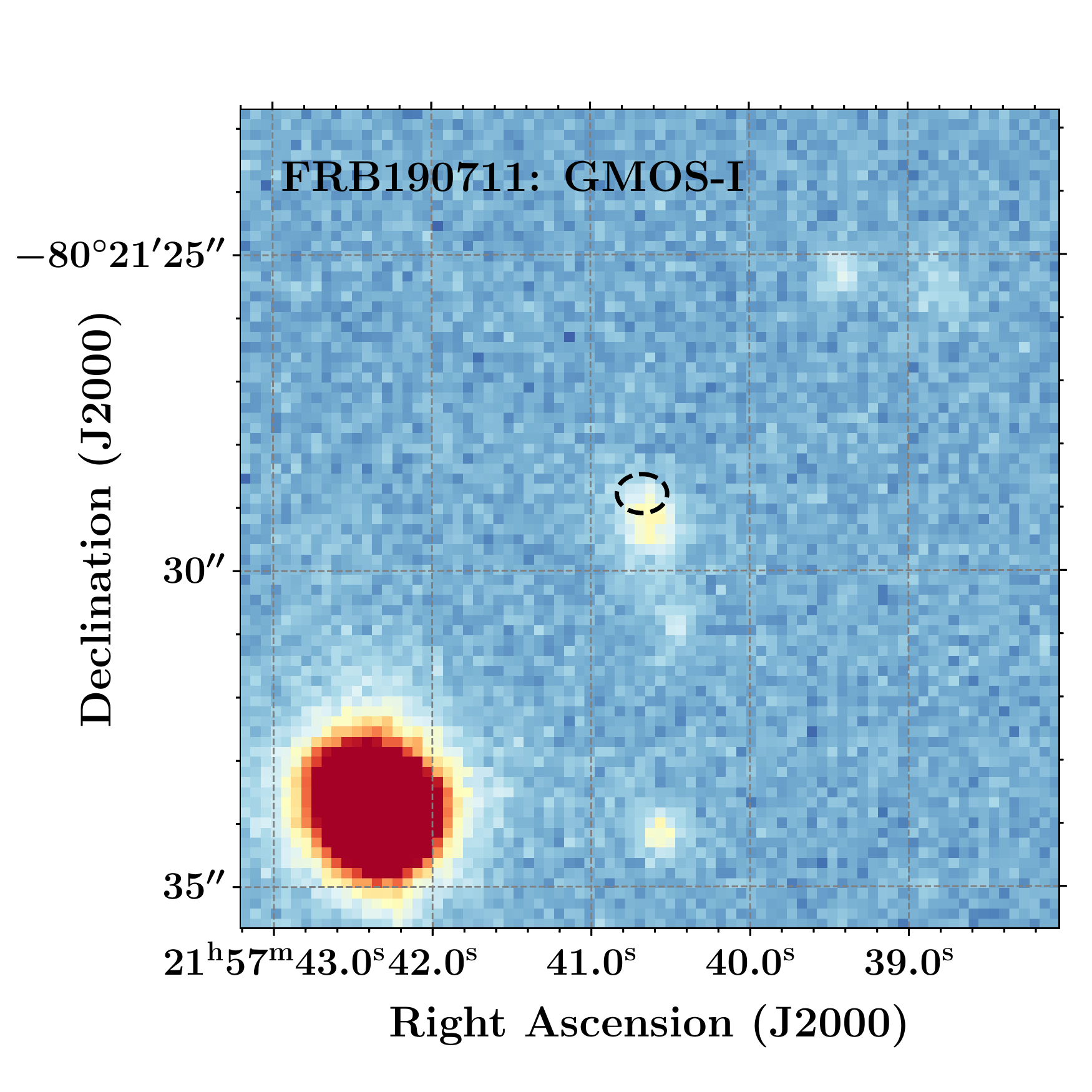}
    \includegraphics[width=5.7cm]{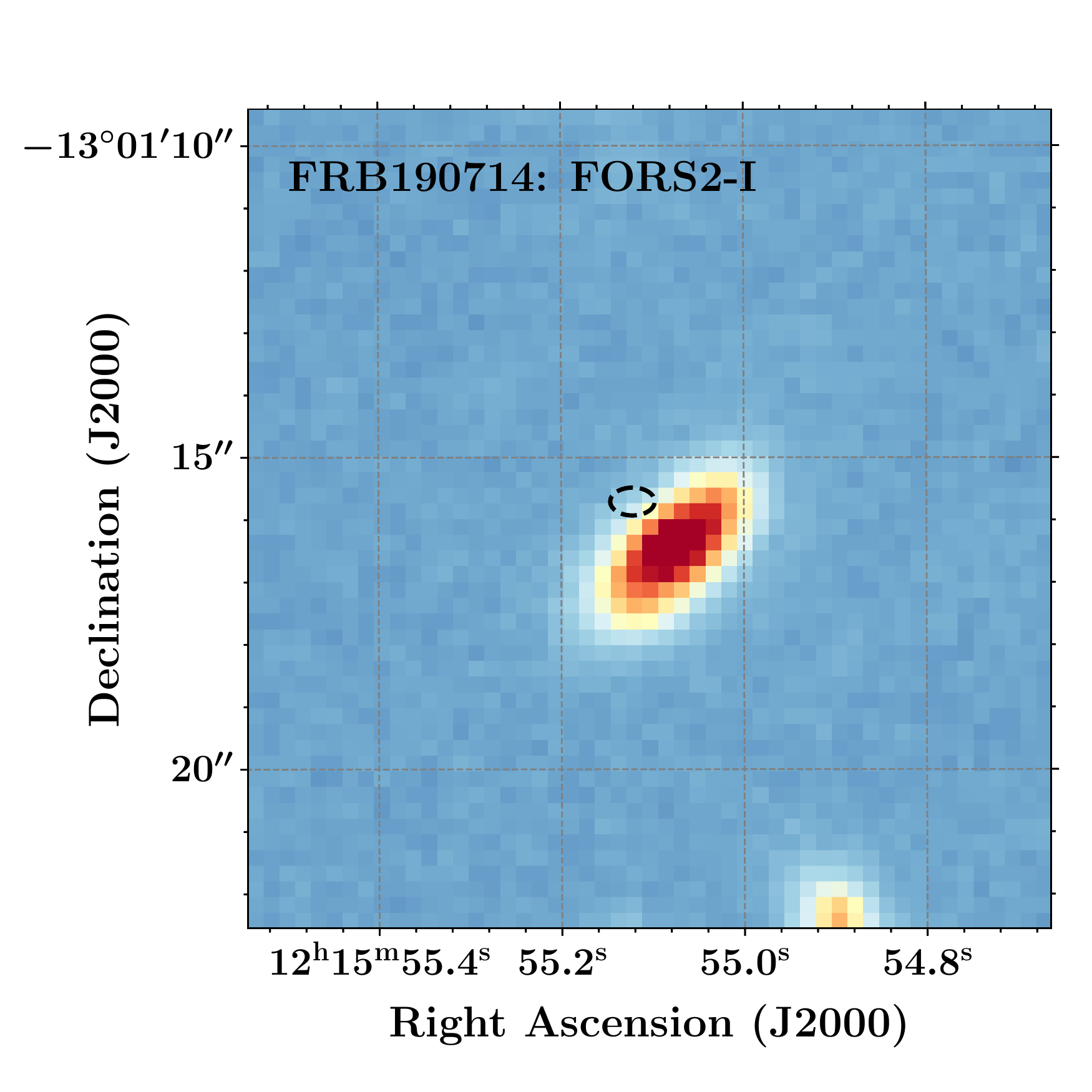}
    \includegraphics[width=5.7cm]{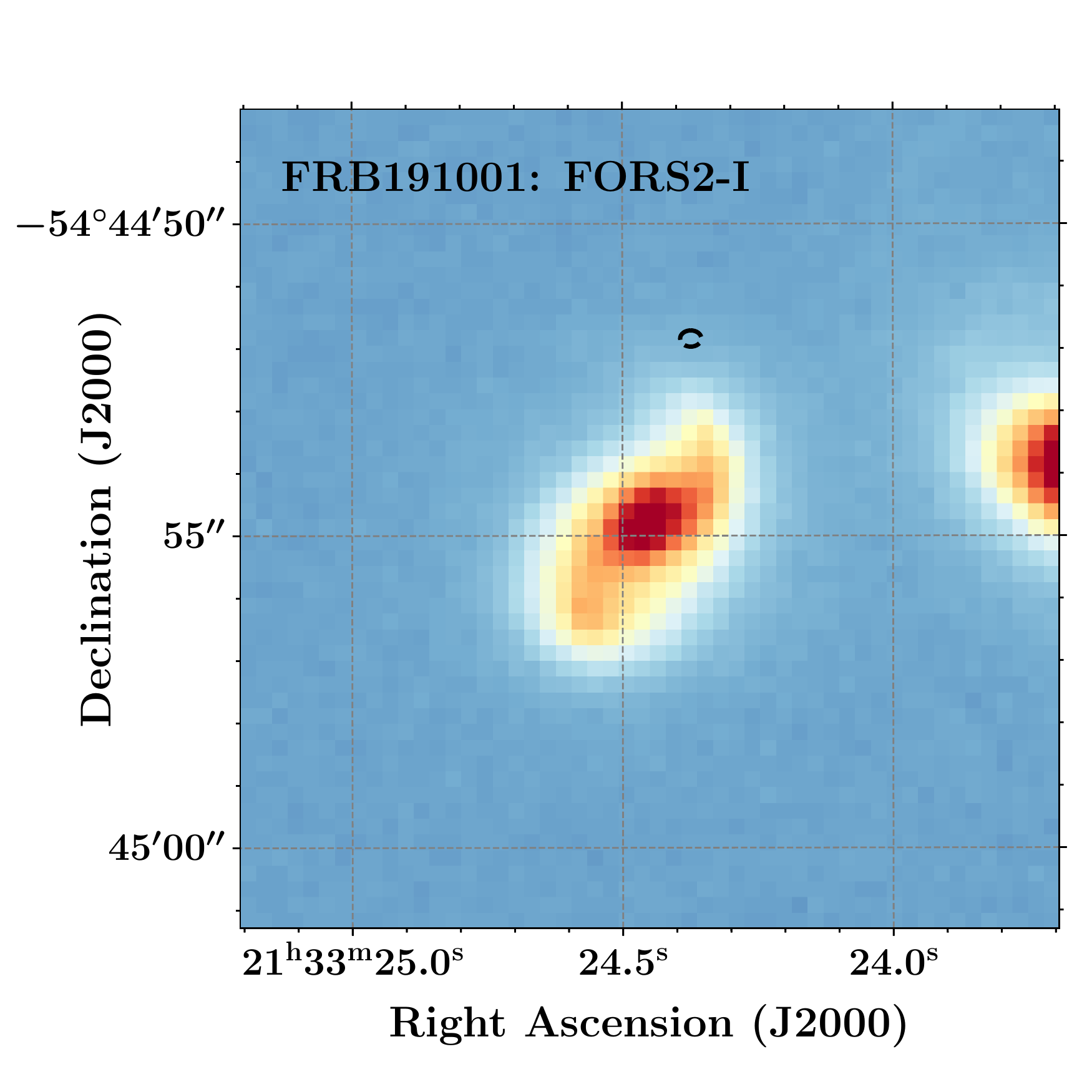}
    \includegraphics[width=5.7cm]{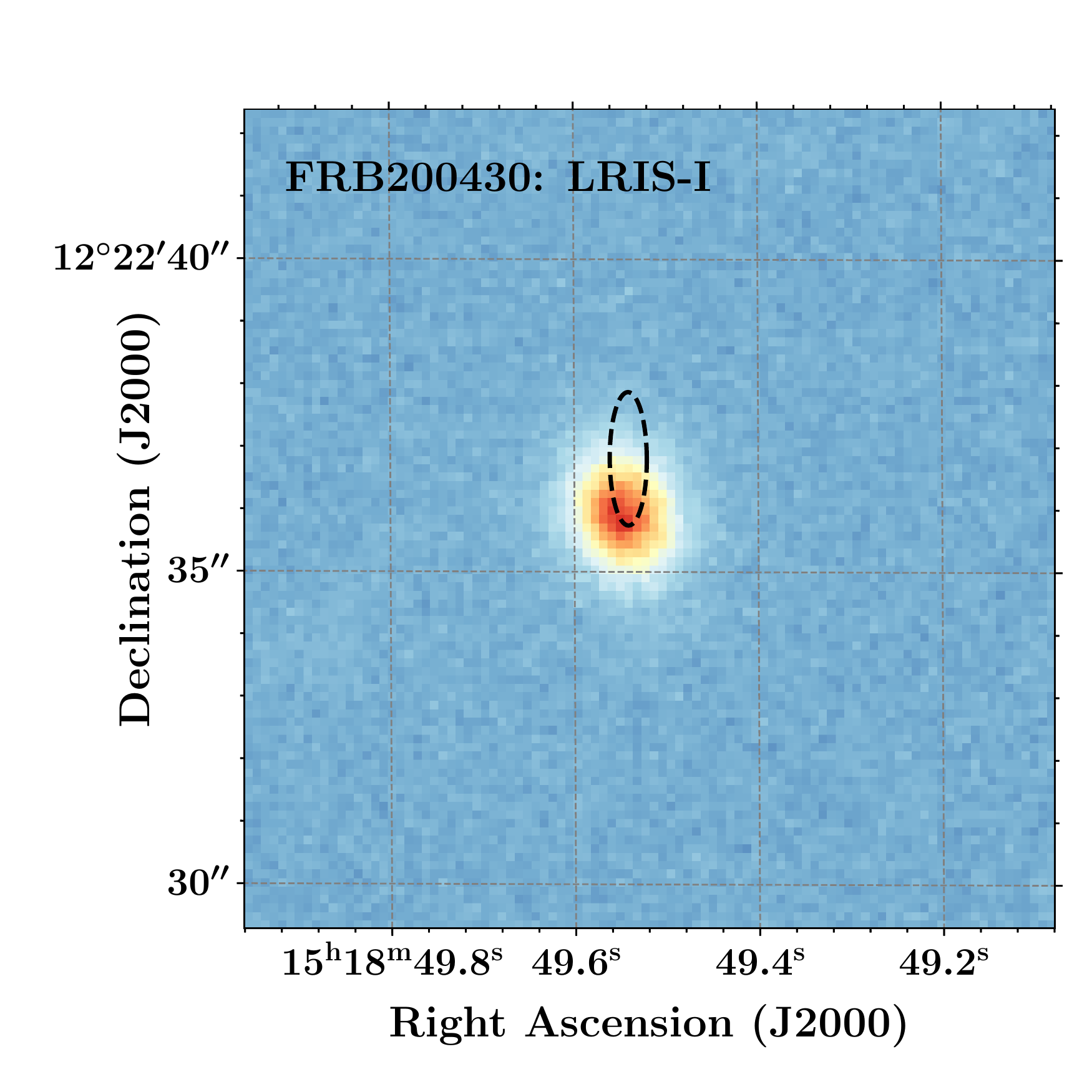}
    \caption{Mosaic showing the $I/i$-band images of the host galaxies of FRBs\,190611, 190711, 190714, 191001, and 200430. The dashed black lines represent the total $1\sigma$ uncertainties on the FRB positions (statistical and systematic). 
    }
	\label{fig:images}
\end{figure*}

The FRB is coincident with an $r \approx 23.5$\,mag galaxy
at $\alpha,\delta =$ \hgsvev\ (see Figure~\ref{fig:images}), with an offset of $0\farcs49$. No clear morphological structures can be identified in the GMOS imaging, and we measure an effective half-light radius of $R_{\rm eff} = 0\farcs46$. We assert a secure association of FRB\,190711 to this galaxy, given the low chance probability of $\mpchance = 0.011$, and include it in Sample~A.

Using \pypeit, we have performed a 2D coaddition of the
VLT X-Shooter spectra presented in \citet{Macquart20}. 
Based on the detection of \hb\ and [\oiii] in this spectrum, we find $\mzspec = 0.5220$. At this redshift, the physical projected offset of the FRB from the galaxy centroid is $\approx 3$\,kpc.
We do not detect H$\alpha$ emission, but this feature lies 
at a lower throughput portion of the spectrograph where there is also significant telluric absorption.


\subsubsection{FRB\,190714}
\label{ssec:190714}

On UT 2019 July 14 at 05:37:12.9, the ASKAP telescope recorded FRB\,190714 at $\alpha,\delta$ = \frbsef\ (J2000), with an uncertainty of $\sigma_{\rm \alpha, \delta} = 0\farcs4, 0\farcs3$.
This localization places FRB\,190714 $\approx 0\farcs5$ from the galaxy \hgsvf\ (see Figure~\ref{fig:images}), which was previously cataloged by the Pan-STARRS \citep{Chambers16}
and the VISTA \citep{Cross12} surveys.
It is a relatively bright source ($r = 20.85$\,mag),
and we estimate a chance association of $\mpchance = 0.005$.
We thus include this galaxy in Sample~A. We do not detect any distinct morphology of the host galaxy in our FORS2 $I$-band image, but there might be evidence of spiral arms based on preliminary results obtained from imaging with the {\it Hubble Space Telescope} \citep[][in preparation]{Manningsprep}. We measure an effective half-light radius of $R_{\rm eff} = 1\farcs02$.

We obtained optical spectroscopy of the host of FRB\,190714 on 2020 January 28 with the LRIS spectrometer \citep{Oke95} on the Keck~I 10m telescope. This dual-camera instrument was configured with the 600/7500 grating, the 600/4000 grism, and a slit mask designed to observe
the FRB-host and additional galaxies in the field.
We reduced these data with \pypeit, and the extracted 1D spectrum was then flux-calibrated through observations
of a spectroscopic photometric standard acquired on the same (clear)
night and scaled to the Pan-STARRS photometry.
The bright nebular emission lines of \hb, [\oiii], 
\ha, and [\nii] yield a spectroscopic redshift of
$\mzspec = 0.2365$.
This places FRB\,190714 at a projected physical separation of $\approx 2$\,kpc from the galaxy center.

\subsubsection{FRB\,191001}
\label{ssec:frb191001}

On UT 2019 October 01 at 16:55:36.0, the ASKAP telescope recorded FRB\,191001 at $\alpha,\delta$ = \frbten\ (J2000), with an uncertainty of $\sigma_{\alpha, \delta} = 0\farcs17, 0\farcs13$ \citep{Bhandari20b}.
This position is $\approx 2\farcs9$ north of the
previously cataloged source DES\hgoo\ \cite[Figure~\ref{fig:images};][]{des-dr1}.
Despite the relatively large angular offset, the bright
magnitude ($r = 18.41$\,mag) yields a chance coincidence probability
of only $\mpchance = 0.003$.
We therefore include this galaxy in Sample~A. The host galaxy of this FRB shows clear spiral-arm features, with the FRB occurring in the outskirts of the northern arm \citep[see][for a more detailed study of this FRB]{Bhandari20b}. The estimated effective half-light radius is $R_{\rm eff} = 1\farcs44$. 

On UT 2019 October 4, we obtained a GMOS spectrum of the host of FRB\,191001 with the
Gemini-S telescope, configured with 
a $1''$ long slit and the R400 grating tilted to cover
$\lambda \approx 5000-9900$\AA\ with a full width at half maximum
(FWHM)~$\approx 500 \, \mkms$.  
The data were reduced with the \pypeit\ software package
(see Section~\ref{ssec:190611} for details)
and flux calibrated with a standard star obtained
and scaled to $r=18.4$\,mag.
The detection of strong nebular emission lines from \hb, [\oiii], 
\ha, and [\nii] yield a spectroscopic redshift of
$\mzspec = 0.2340$. This places FRB\,191001 at a projected physical separation of $\approx 11$\,kpc from the galaxy center.

The longslit was oriented at PA=100$^\circ$ to include
the neighboring galaxy \hgBoo, which lies
$\approx 7''$ east of the identified host galaxy.
Its spectrum also shows strong nebular emission
yielding $\mzspec = 0.2339$, i.e., coincident with
the host of FRB\,191001, revealing a physical pair.
At a projected separation of $\approx 25$\,kpc, 
we expect that these galaxies are in the process of merging.


\subsubsection{FRB\,200430}
\label{ssec:frb200430}

On UT 2020 April 30 at 15:49:48.3, the ASKAP telescope recorded FRB\,200430 at $\alpha,\delta$ = \frbtut\ (J2000), with an uncertainty of $\sigma_{\alpha, \delta} = 0\farcs3, 1\farcs1$. This is $\approx 1\farcs0$ north of the previously cataloged galaxy \hgtut\ in the Pan-STARRS catalog \citep{Chambers16} with $r=21.51$\,mag. We obtained additional $g$- and $I$-band imaging with Keck/LRIS on UT 2020 June 21, the latter shown in Figure~\ref{fig:images}. 
Based on the offset and the host-galaxy magnitude, we derive a chance coincidence probability of $\mpchance = 0.005$.
We therefore include this galaxy in Sample~A. We do not detect any distinct morphology of the host galaxy based on the deeper Keck images and measure an effective half-light radius of $R_{\rm eff} = 0\farcs57$.

On UT 2020 May 16, we obtained optical spectroscopy of the identified host galaxy with the Alhambra Faint Object Spectrograph and Camera (ALFOSC) mounted at the Nordic Optical Telescope (NOT). The spectra were obtained with grism 4 (covering 3200 -- 9600\,\AA) and a slit width of $1\farcs3$. The observations were performed under good conditions with an average seeing of $1\farcs1$ at an airmass around 1.2 during the integration. The data were reduced with the \pypeit\ software package. 
We determine a redshift for the host galaxy of $z=0.160$ based on the detection of the \ha, [\oii], and [\sii] emission features and the H and K absorption lines from Ca. This places FRB\,200430 at a projected physical separation of $\approx 3$\,kpc from the galaxy center.

\subsection{Literature Compilation}
\label{sec:sub_literature}

In addition to the five new FRB hosts presented here, we include all other (currently) known FRB-host galaxies in our analysis. These include FRBs 121102 \citep{Chatterjee17,Tendulkar17,Bassa17}, 180916 \citep{Marcote20}, 180924 \citep{Bannister19,Bhandari20a}, 181112 \citep{Prochaska19b}, 190102 \citep{Macquart20,Bhandari20a}, 190523 \citep{Ravi19}, 190608 \citep{Macquart20,Bhandari20a,Chittidi20}, and 20190614D \citep[hereafter referred to as FRB\,190614;][]{Law20}. 

In Appendix~\ref{sec:literature} we briefly describe these additional galaxies associated with well-localized FRBs and any new observations obtained after the primary publications. We separate them primarily by FRB survey. For the hosts previously reported by \citet{Bhandari20a}, we simply include their reported measurements here. For the host galaxy of FRB\,180916 \citep{Marcote20}, we extract the photometry from the SDSS and the Wide-field Infrared Survey Explorer (WISE) catalogs to obtain a more precise estimate of the stellar mass. We find the best-fit value to be approximately a factor of five lower than the stellar mass reported by \cite{Marcote20}. We also obtained independent spectra of the putative host galaxy of FRB\,190523 \citep{Ravi19,Prochaska19c},
allowing us to derive an upper limit on the line flux of \hb,\ which we use to place a stronger limit on the star-formation rate (SFR) of $<0.09\,M_{\odot}\,{\rm yr}^{-1}$. Finally, we used the photometry reported by \citet{Bassa17} for the host galaxy of FRB\,121102 to model our own spectral energy distribution (SED; see Section~\ref{ssec:sed}) for consistency with the rest of the sample.

\subsection{Overall Sample Properties}
\label{ssec:sampleprop}

Our overall parent sample consists of \ntot\ FRB-host galaxies as presented in Table~\ref{tab:sample}. Out of these \ntot\ hosts, \nsampA\ satisfy the Sample~A criteria. These include all the 5 new host galaxies characterized in detailed here (FRBs\,190611, 190711, 190714, 191001, and 200430) and the hosts of FRBs\,121102, 180916, 180924, 190102, and 190608 (i.e. all the hosts of the repeating FRBs are also in Sample~A). We have placed the host of FRB\,190523 discovered by \citet{Ravi19} into Sample~C because the poor FRB localization makes the host-galaxy association less secure. The host galaxy of FRB\,181112 identified by \citet{Prochaska19b} is also placed in Sample~C because the proposed foreground galaxy has a similarly low ($\mpchance < \achance$) chance association probability. The observed dispersion measure of $\mdmfrb = 589\, \mdmunits$ for this event, however, supports the association that the background galaxy is the host of FRB\,181112. To be conservative and consistent with the other sample classifications, we rely only on the statistical properties of the FRB-host associations here to avoid biasing the host identifications. 
For FRB\,190614, \citet[][]{Law20} identified two potential host-galaxy candidates, for which only photometric redshifts have been obtained, placing it in Sample~D.

This more than doubles the number of ASKAP-detected FRB hosts studied in our previous work \citep{Bhandari20a}. The FRBs are distributed throughout the celestial sphere, and our full sample spans redshifts of $z_{\rm FRB} = 0.03 - 0.66$. 
We wish to caution, however, that because the number of FRB-host identifications is still small, we consider all known FRB hosts here regardless of their initial selection. A more careful homogeneous selection is required when a larger number of FRBs with subarcsecond localizations and their associated host galaxies have been properly identified.


\begin{figure*}[!t]
\centering
    \includegraphics[width=8.7cm]{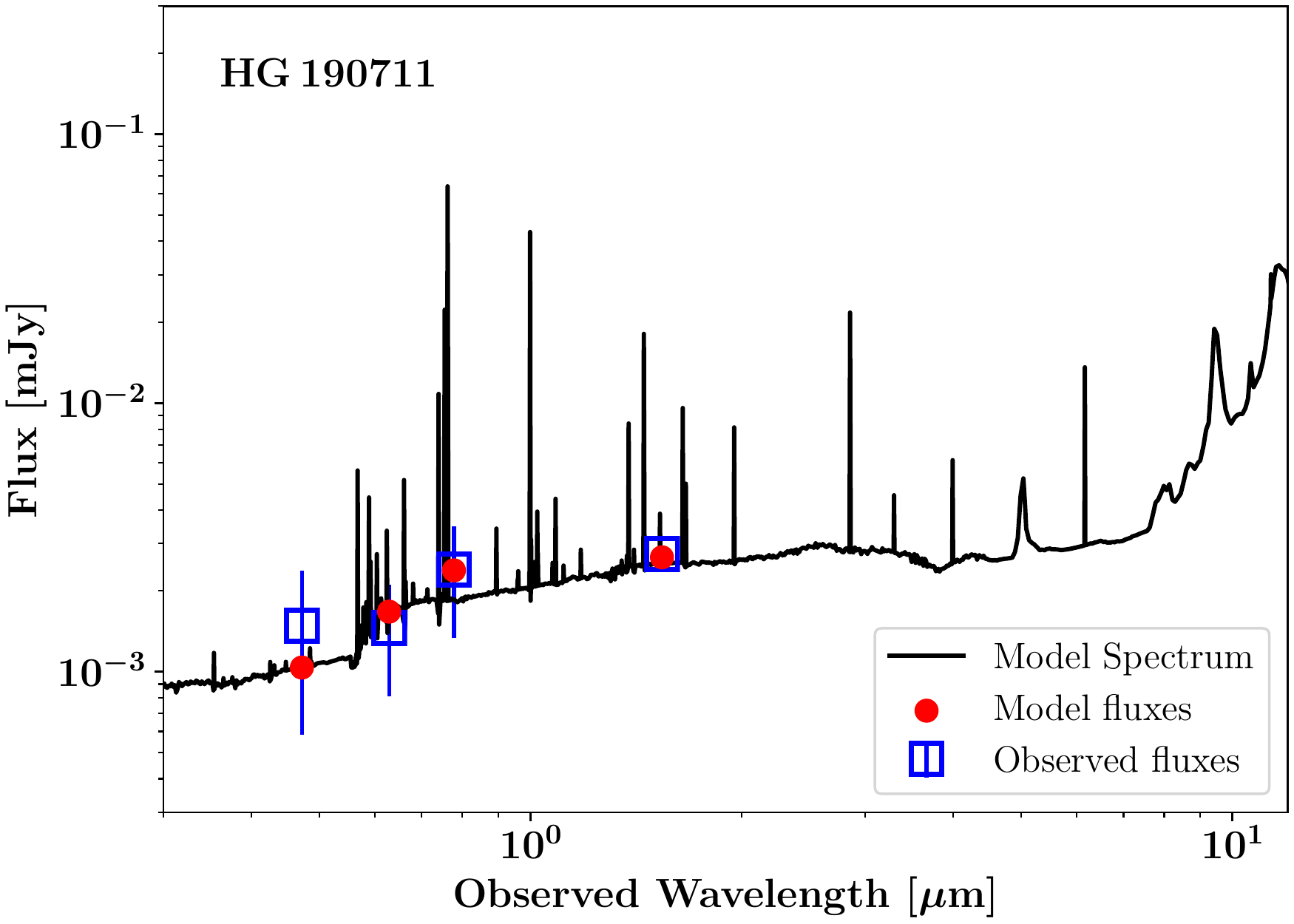}
    \includegraphics[width=8.7cm]{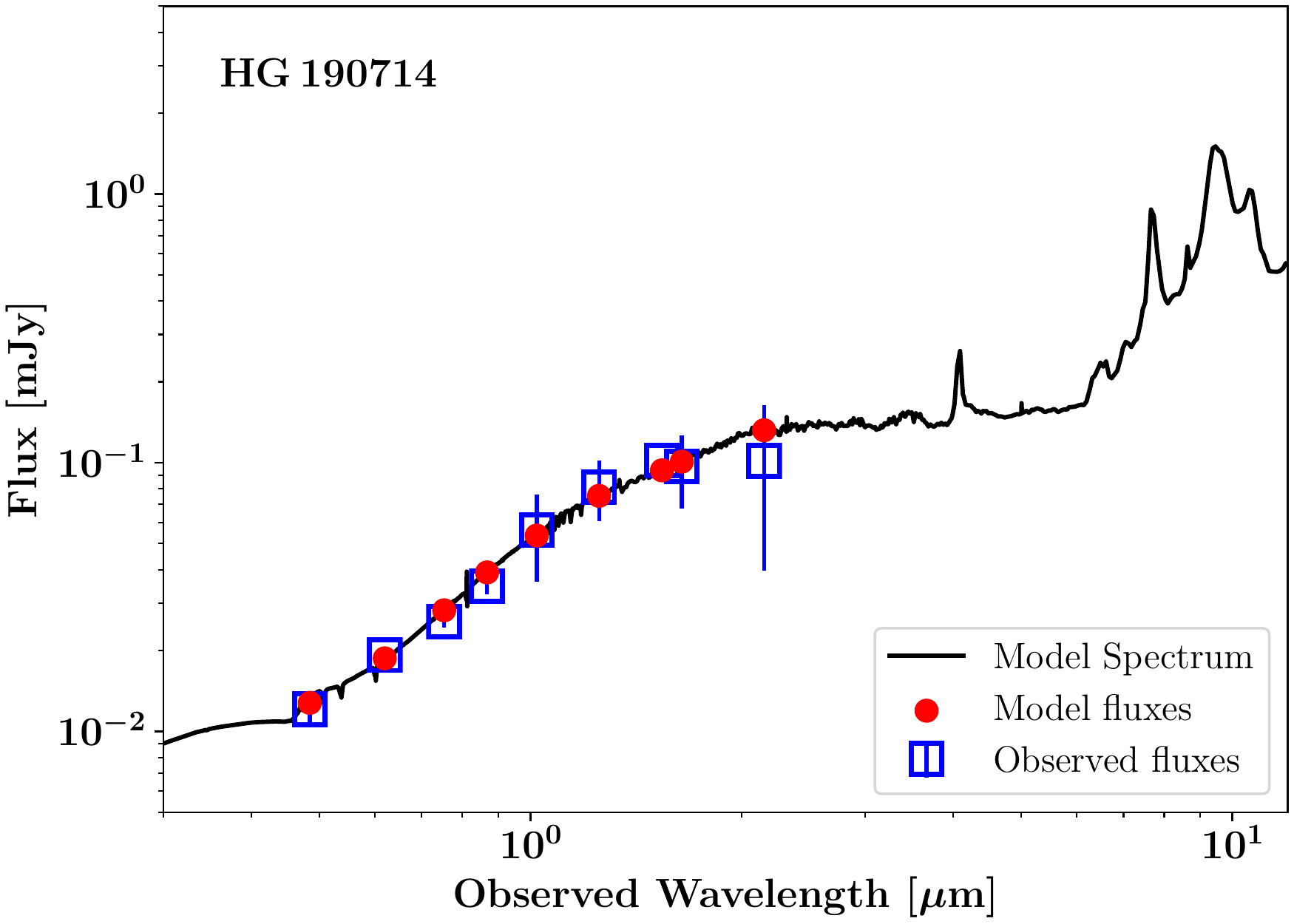}
    \includegraphics[width=8.7cm]{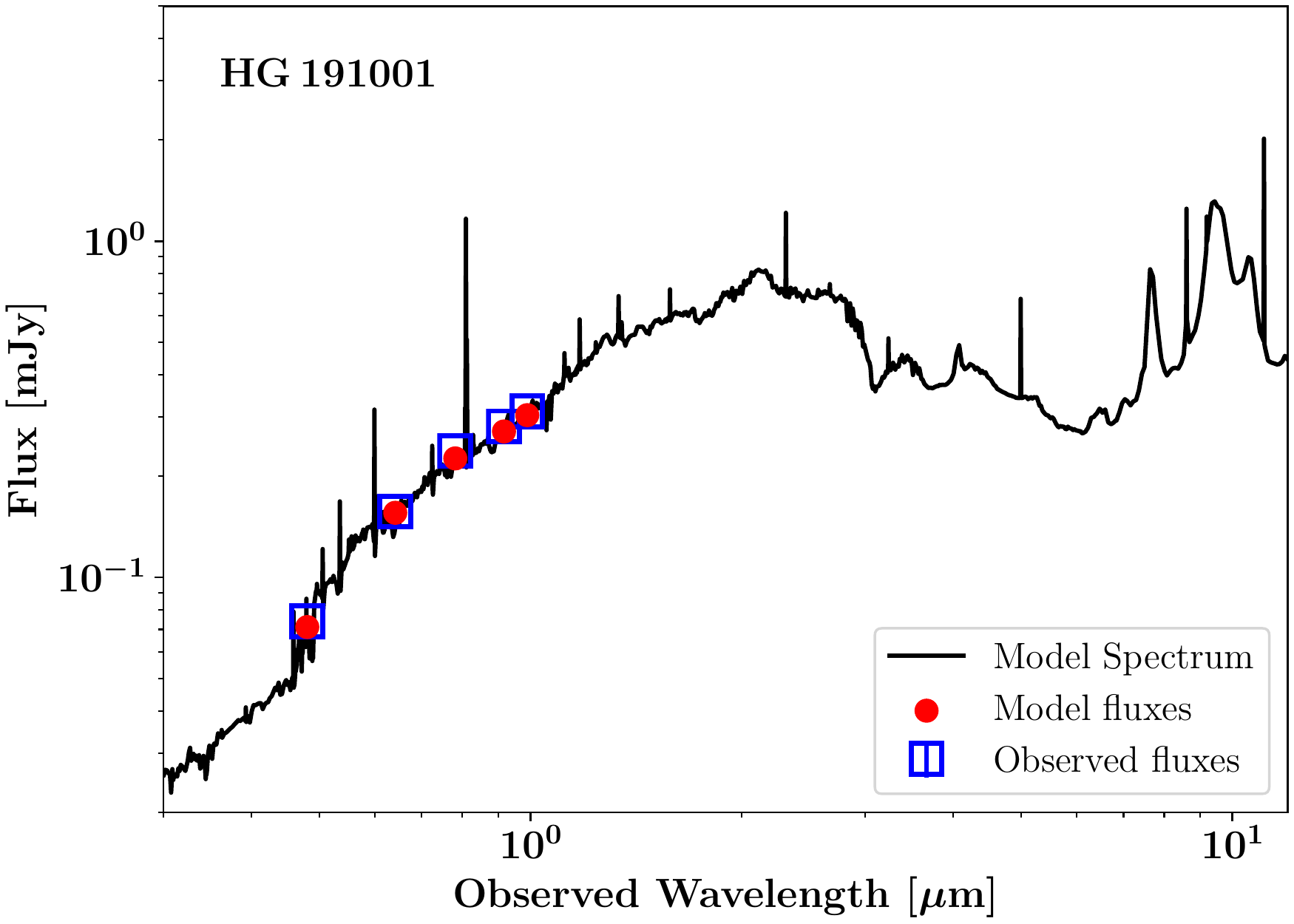}
    \includegraphics[width=8.7cm]{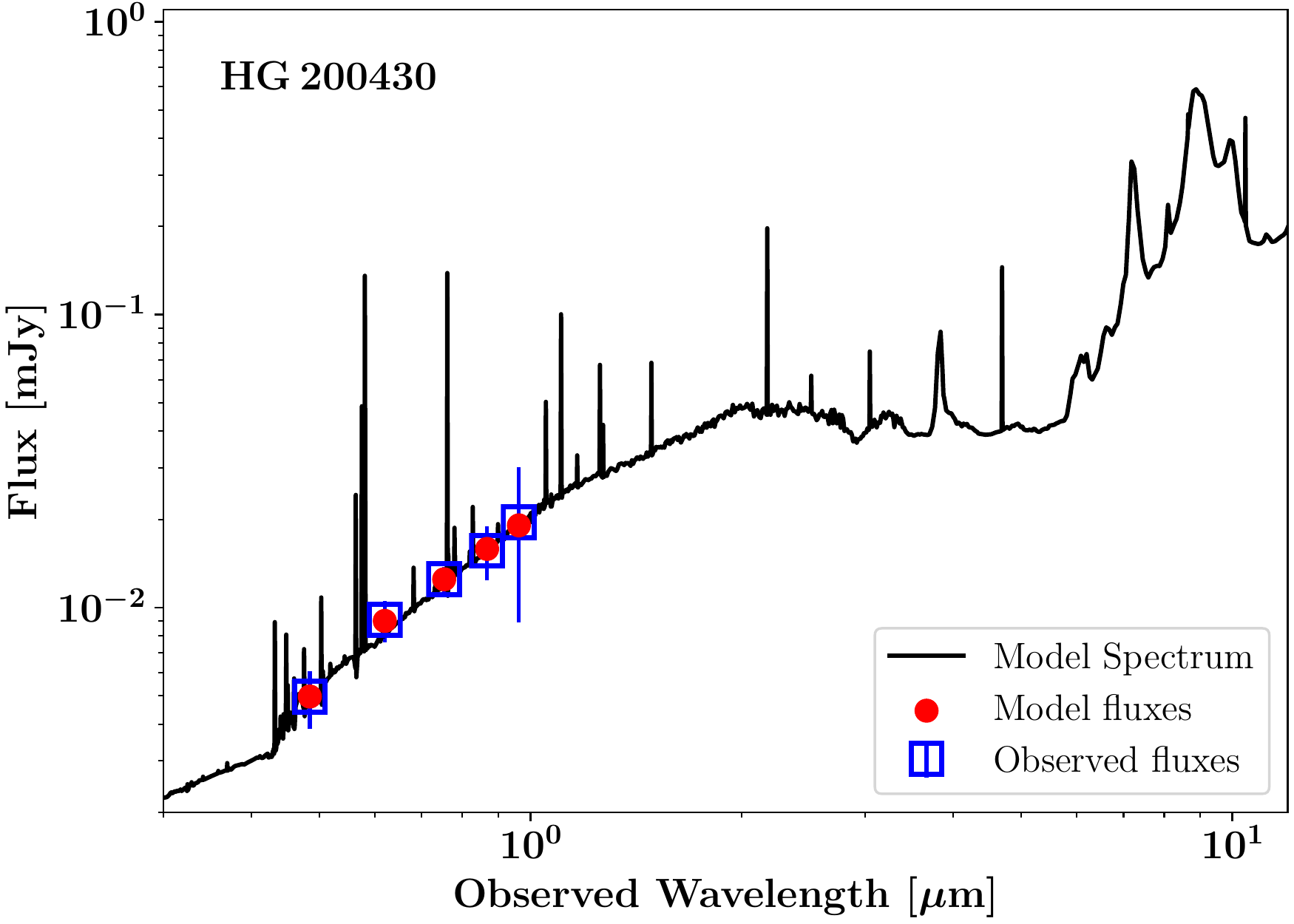}
    \caption{SED models for the host galaxies of FRBs\,190711, 190714, 191001, and 200430 (named with an HG prefix). FRB\,190611 is not shown here because we do not have sufficient photometric data to construct the SED of the host. The best-fit SED models from {\tt CIGALE} are shown as solid black lines, the observed magnitudes (corrected for Galactic extinction and converted into fluxes) as blue squares, and the model fluxes as red dots. In all models, the redshift has been fixed to $\mzspec$.}
	\label{fig:cigale}
\end{figure*}


\subsection{Repeating and Nonrepeating FRBs}

Throughout the paper, we distinguish the hosts of the three FRBs that are currently known to repeat (FRBs\,121102, 180916, and 190711) from the hosts of the other apparently nonrepeating one-off FRBs. Repeating FRBs by definition cannot be cataclysmic events, whereas apparently nonrepeating FRBs might be. In principle, all FRBs could be found to repeat if observed on long enough time scales and with an appropriate cadence, but it appears unlikely that they do \citep[at least similarly to FRB\,121102;][]{James20}. The apparently longer intrinsic temporal pulse width for repeating FRBs compared as an ensemble to as yet nonrepeating FRBs \citep{CHIMErep,Fonseca20} also suggests that repeating sources show different pulse morphologies than nonrepeating sources, including wider burst envelopes and distinct time-frequency drifting. This could imply a different emission mechanism for repeating and one-off sources. We caution that it is possible that FRBs that are currently classified as nonrepeating may exhibit repeat pulses in the future, which would change their classification here. As noted by \citet{Day20} and others, signposts of probable repetition can be discerned from high time and frequency resolution analyses of FRB detections, although it appears to reflect a continuum of spectro-temporal polarimetric properties of FRBs more. Here, we reserve the repeater label for events with multiple confirmed bursts. 

Even with a sample of three known repeating FRBs, it is possible to examine the physical properties of their host galaxies compared to the sample of hosts of apparently nonrepeating FRBs. This may provide additional clues on whether two populations of FRBs exist.

\section{Analysis and Results}
\label{sec:analysis}

\subsection{Stellar Population Modeling} \label{ssec:sed}

Following our previous studies of FRB-host galaxies
\citep{Bannister19,Prochaska19b,Bhandari20a,Chittidi20},
we have analyzed the existing photometry
and spectroscopy of all hosts with the {\tt pPXF} \citep{ppxf}
and {\tt CIGALE} \citep{cigale} software packages.
Each package fits a set of stellar population models
and a star formation history (SFH) to the spectra
({\tt pPXF}) or photometry ({\tt CIGALE}) and generates best estimates
for quantities such as stellar mass $M_{\star}$, 
internal extinction \ebv, and age of the main stellar population.  


\begin{figure*}[!ht]
\centering
    \includegraphics[width=5.9cm]{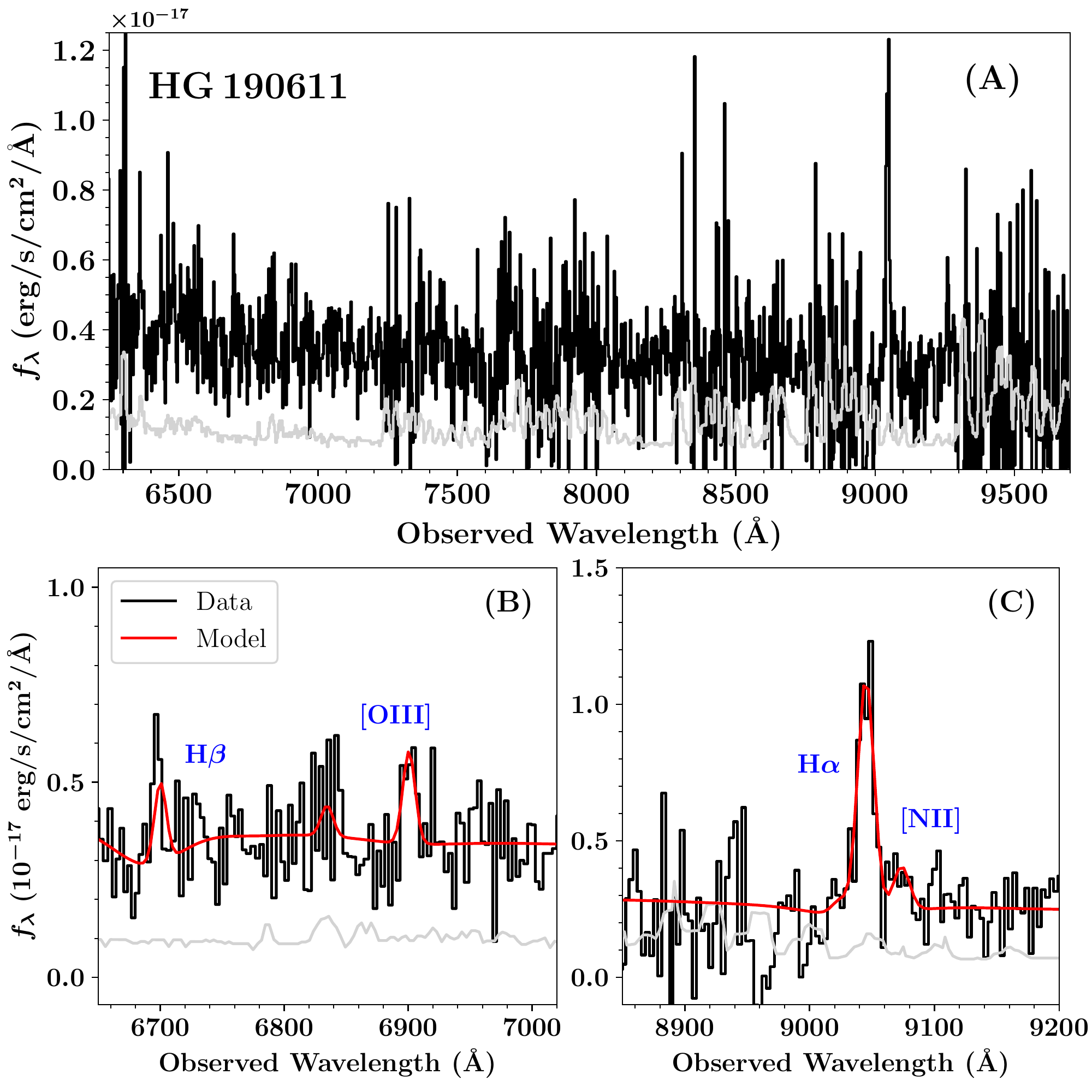}
    \includegraphics[width=5.9cm]{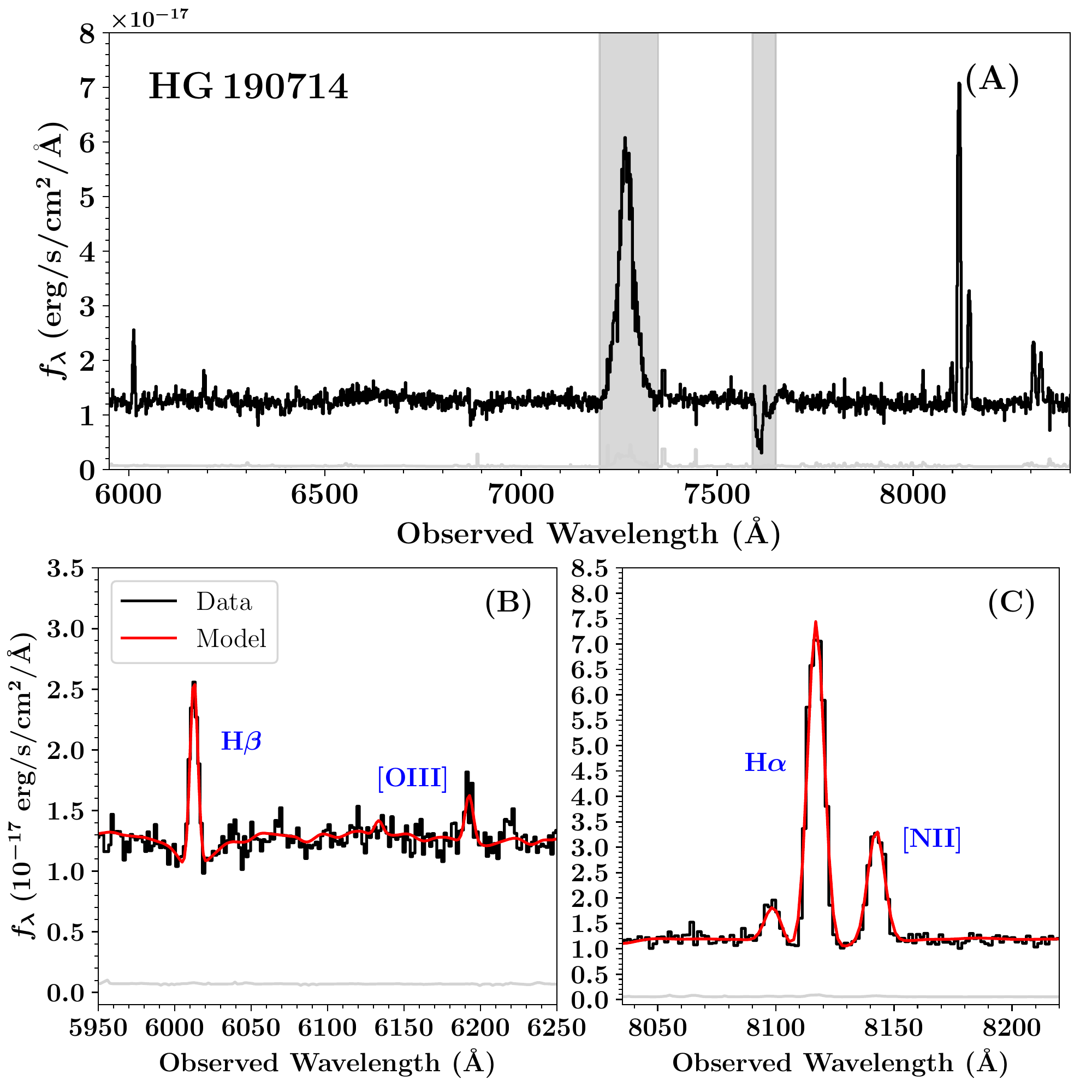}
    \includegraphics[width=5.9cm]{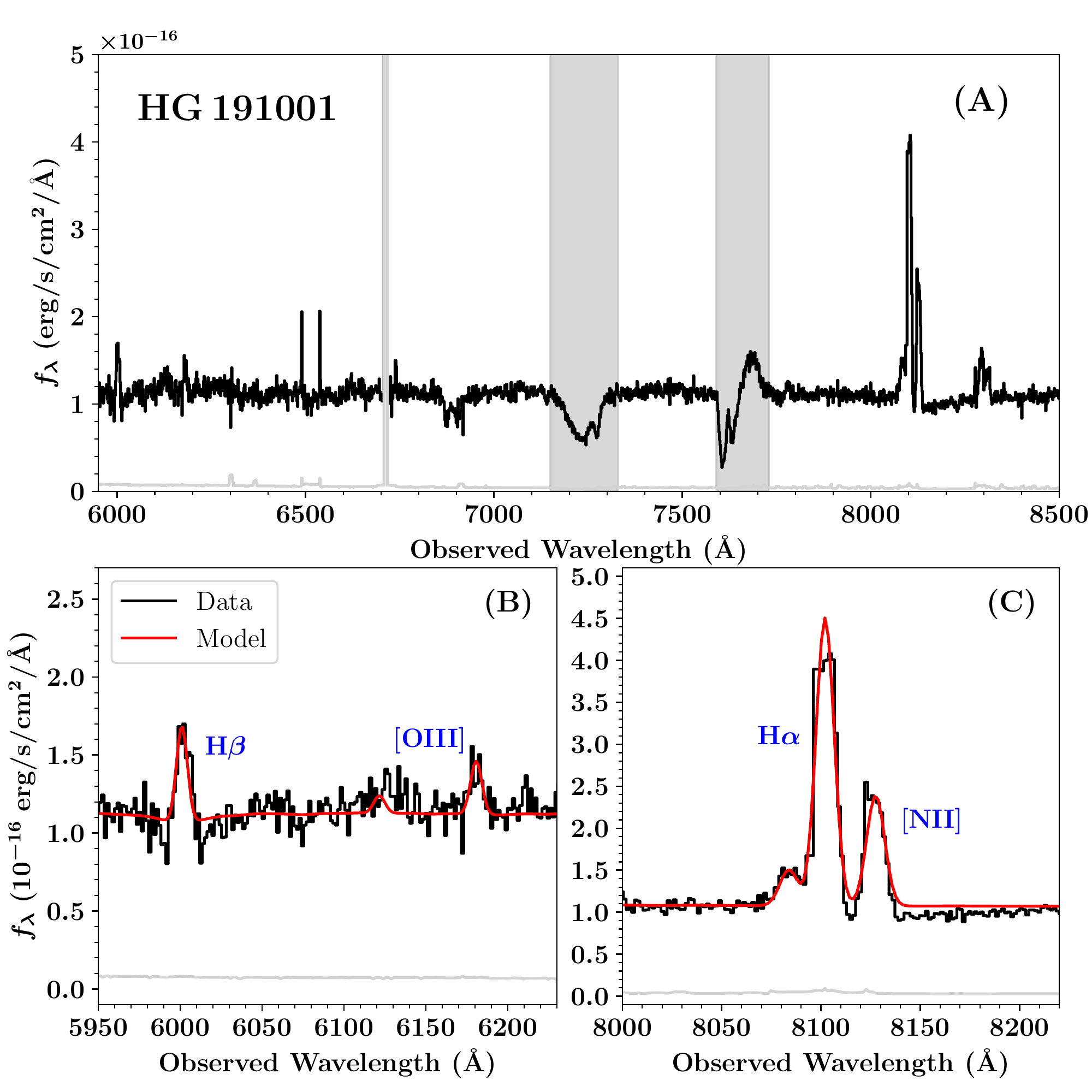}
    \caption{Spectra of the host galaxies of FRBs\,190611, 190714, and 191001 (named with an HG prefix). FRBs\,190711 and 200430 are not shown here because the poor signal-to-noise ratio (S/N) of the spectra does not allow us to model the line features with {\tt pPXF}.
    The solid black lines show the dust-corrected spectra, with the associated error spectrum shown in gray. In the bottom panels, we show zoom-ins on the most prominent nebular emission lines from \ha, \hb, [\oiii], and [\nii], with the best-fit models from {\tt pPXF} overplotted on the spectra shown as solid red lines. Strong telluric regions are masked out by the gray shaded regions.
    }
	\label{fig:ppxf}
\end{figure*}


We adopt the following assumptions
in these analyses:

\begin{itemize}
    \item A delayed-exponential SFH model with no late-burst population (${\rm SFR}(t) \propto t/\tau^2 \times \exp{(-t/\tau)} $). Here $t$ is the age, with $t=0$ denoting the onset of star-formation, and $\tau$ is the $e$-folding time of the decaying part of the SFH.
    \item The \cite{Bruzual03} simple stellar population with the initial mass function (IMF) from \cite{Chabrier03} and a metallicity allowed to vary from 0.005$Z_\odot$ to 2.55$Z_\odot$.
    \item The \cite{Calzetti00} dust extinction 
    model, modified following \cite{LoFaro17}. 
    \item The \cite{Dale14} dust emission model with an AGN fraction  $f_{\rm AGN} \le 0.2$ and
    a power-law exponent of 2.
\end{itemize}
We determine the internal host-galaxy extinction \ebv$_{\rm host}$ from the SED modeling without adopting the visual extinction derived from the Balmer decrement as input, but note that the two independent estimates are generally consistent.
In all cases, we have input observations corrected 
for Galactic extinction using the \ebv$_{\rm Gal}$\ values
derived from \cite{Schlafly11} and the 
\cite{Fitzpatrick07} extinction law with
$R_V = 3.1$ implemented in the {\tt extinction}\footnote{\url{https://extinction.readthedocs.io/en/latest}} software package. The derived photometry for all FRB hosts considered here is provided in Appendix~\ref{sec:phot}, in Tables \ref{tab:photom1}--\ref{tab:photom3}. 
The precise input parameter file for CIGALE
is available in the {\tt cigale.py} module of the
FRB repository on GitHub.

The best-fit models for the host galaxies described in Section~\ref{ssec:new_frbs}
are presented in Figures~\ref{fig:cigale} and
\ref{fig:ppxf}.
These include fluxes for common emission lines
derived from the {\tt pPXF} analysis (provided in Table~\ref{tab:lineflux}), which corrects for Balmer absorption. 
For uniformity, we also performed the
same analysis on FRB hosts drawn from the literature,
especially for all galaxies in Sample~A.
This includes reanalyses of hosts from our
own previous publications \citep[e.g.][]{Bhandari20a}.
Any substantial differences from previously reported
estimates are described in Appendix~\ref{sec:literature}. 
The results are summarized in Table~\ref{tab:hostprop}.


\input{tab_emlineflux}


\subsection{Star-Formation Rate}

We derive the SFR for each FRB host by first computing the dust-corrected \ha\ line fluxes using the $A_V$ derived from the Balmer decrement to obtain the intrinsic \ha\ line luminosities as $L_{\rm H\alpha} ({\rm erg\,s^{-1}}) = F_{\rm H\alpha}({\rm erg\,s^{-1} cm^{-2}}) \times 10^{A(\lambda)/2.5} \times (4\,\pi\,d_L^2)$, with $d_L$ the luminosity distance. 

This we translate into an SFR via the conversion factor 
\begin{equation}\label{eq:sfr}
    {\rm SFR}\,(M_{\odot}\,{\rm yr}^{-1}) = 4.98\times 10^{-42} \, L_{\rm H\alpha}({\rm erg\,s^{-1}})
\end{equation}
following \citet{Kennicutt98}, but adopting the IMF from \cite{Chabrier03}\footnote{Assuming a conversion from the Salpeter-determined SFR of SFR$_{\rm Chab}$ = SFR$_{\rm Salp}\times 0.63$.}. We report the uncertainties on the SFR estimates including the scatter in the SFR-$L_{\rm H\alpha}$ relation ($\approx 30\%$). For the three FRB hosts where the \ha\ line flux has not been measured, we derive the SFR from \hb\ assuming the nominal relative strength compared to \ha\ (FRB\,190523 in Sample~C and FRB\,190711 in Sample~A) or from the best-fit SED model from {\tt CIGALE} (for FRB\,200430, Sample~A). We find that the overall sample of FRB hosts are characterized by a large range in SFR, spanning $0.05 - 10$\,$M_{\odot}$\,yr$^{-1}$. For the host of FRB\,190614, no constraints could be placed on the SFR because the nature of the host galaxy and redshift are uncertain \citep[Sample~D;][]{Law20}.

\subsection{Gas-phase Metallicity} 
\label{ssec:met}

To infer the gas-phase metallicities of the FRB hosts, we rely on commonly used diagnostic ratios of strong nebular emission lines \citep[see][for a recent review]{Maiolino19}. These allow us to compute the oxygen abundances $12+\log{\rm (O/H)}$ for each host galaxy. The strong-line diagnostics are calibrated to more direct methods, relying on measurements of the electron temperature $T_e$ or derived from photoionization models. In the following, we adopt the O3N2 calibration from \cite{Hirschauer18}, which parameterizes the oxygen abundance as
\begin{equation}
\begin{split}
    12+\log{\rm (O/H)} = 8.987 - 0.297\times{\rm O3N2} \\ - 0.0592\times{\rm O3N2}^2 - 0.0090\times{\rm O3N2}^3 \;,
\end{split}
\end{equation}
where O3N2 = log[([\oiii]$\lambda 5007$/\hb)/([\nii]$\lambda 6584$/\ha)]. This calibration has been shown to be consistent with more direct $T_e$-based methods, and has an rms uncertainty of 0.111\,dex. The majority of the FRB hosts are relatively metal rich, with oxygen abundances distributed between $12+\log {\rm (O/H)} = 8.7 - 9.0$. For the host of FRB\,121102, we could only place an upper limit on the oxygen abundance of $12+\log$(O/H) $< 8.08$ because of the non-detection of [\nii]$\lambda 6584$. For the hosts of FRBs\,190523, 190614, 190711, and 200430, too few nebular lines have been detected to determine their metallicity.

We caution that because the oxygen abundances derived using the O3N2 calibration are specifically calibrated to SF galaxies, the actual metallicities might be slightly different if the emission-line ratios do not represent typical SF galaxies \citep[as reported for FRB hosts by][see also Section~\ref{ssec:bpt}]{Bhandari20a}. However, because the adopted calibration takes both the [\nii]/\ha\ and the [\oiii]/\hb\ ratios into account, the line flux excess of the two ratios should effectively cancel out.


\section{Physical Properties of the FRB-host Population}
\label{sec:hostprop}

The physical properties of all the FRB hosts in our sample are summarized in Table~\ref{tab:hostprop}. In the following analysis, we examine the FRB-host galaxy environments and place them into context of field-selected galaxies. Throughout, we separate the hosts of repeating and seemingly nonrepeating, one-off FRBs. We only consider the 10 FRB hosts in Sample~A for the statistical analyses.


\input{tab_hostprop}


\begin{figure}[!t]
\centering
    \includegraphics[width=8.7cm]{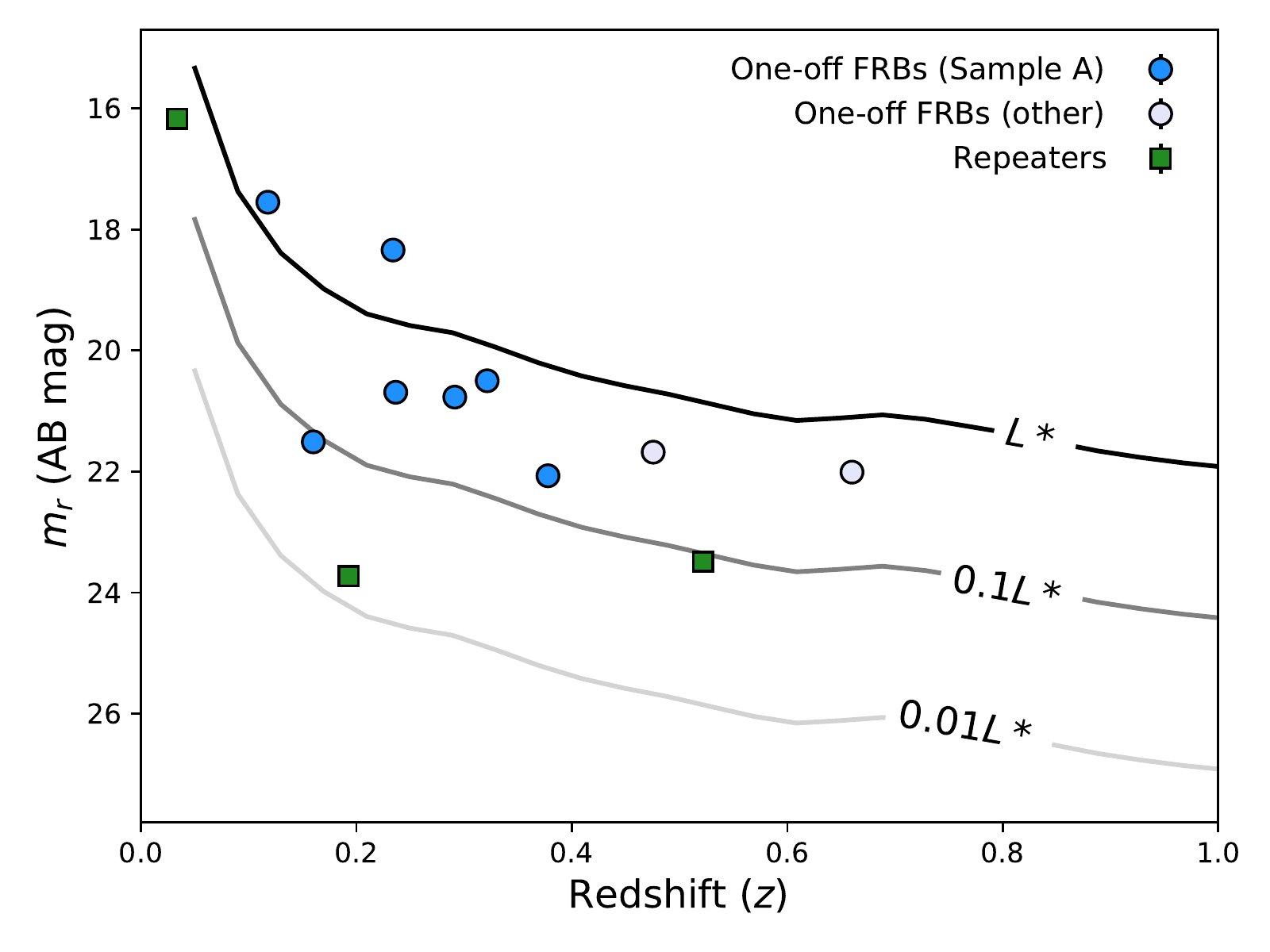}
    \caption{Observed apparent $r$-band magnitude as a function of redshift for the host galaxies of repeating and nonrepeating FRBs. The nonrepeating FRBs are marked by blue (Sample A) and gray (other samples) dots, and repeating FRBs (all in Sample~A) are denoted by green squares. For FRB\,190102 we plot the $I$-band magnitude, since the $r$-band is not available. For comparison, we show constant luminosity tracks of the underlying field galaxy population at $L = 0.01L^*, 0.1L^*$, and $L^*$, which were constructed using the appropriate rest-frame band galaxy luminosity function that corresponds to the observed $r$ band at each redshift. All FRB hosts are luminous with $L > 0.1L^*$, except for that of the repeater, FRB\,121102, which has a luminosity of $L\sim 0.01L^*$.
    }
	\label{fig:zlum}
\end{figure}


\subsection{Luminosity and Color} \label{ssec:colmag}

To place the FRB hosts in the context of galaxies at similar redshifts, we present the apparent $r$-band magnitudes $m_r$ of the FRB hosts as a function of redshift in Figure~\ref{fig:zlum}. We compare the values of $m_r$ to the characteristic luminosity $L^*$ across redshift, using available galaxy luminosity functions \citep{Brown2001,Wolf2003,Willmer2006,ReddySteidel2009,Finkelstein2015}. For each redshift, we adopt the value of $L^*$ in the rest-frame band that corresponds to the observed $r$ band. Interpolation across redshift results in smooth contours corresponding to the luminosity tracks of field-selected galaxies at $L = 0.01L^*, 0.1L^*$, and $L^*$. We find that the majority of the FRB hosts are in the intermediate region between $L \sim 0.1L^* - L^*$ compared to the underlying galaxy population at $0.0 < z < 0.7$. The only exception is the host galaxy of FRB\,121102 \citep{Tendulkar17}, which has a luminosity of $L\sim 0.01L^*$.

\begin{figure}[!t]
\centering
    \includegraphics[width=8.5cm]{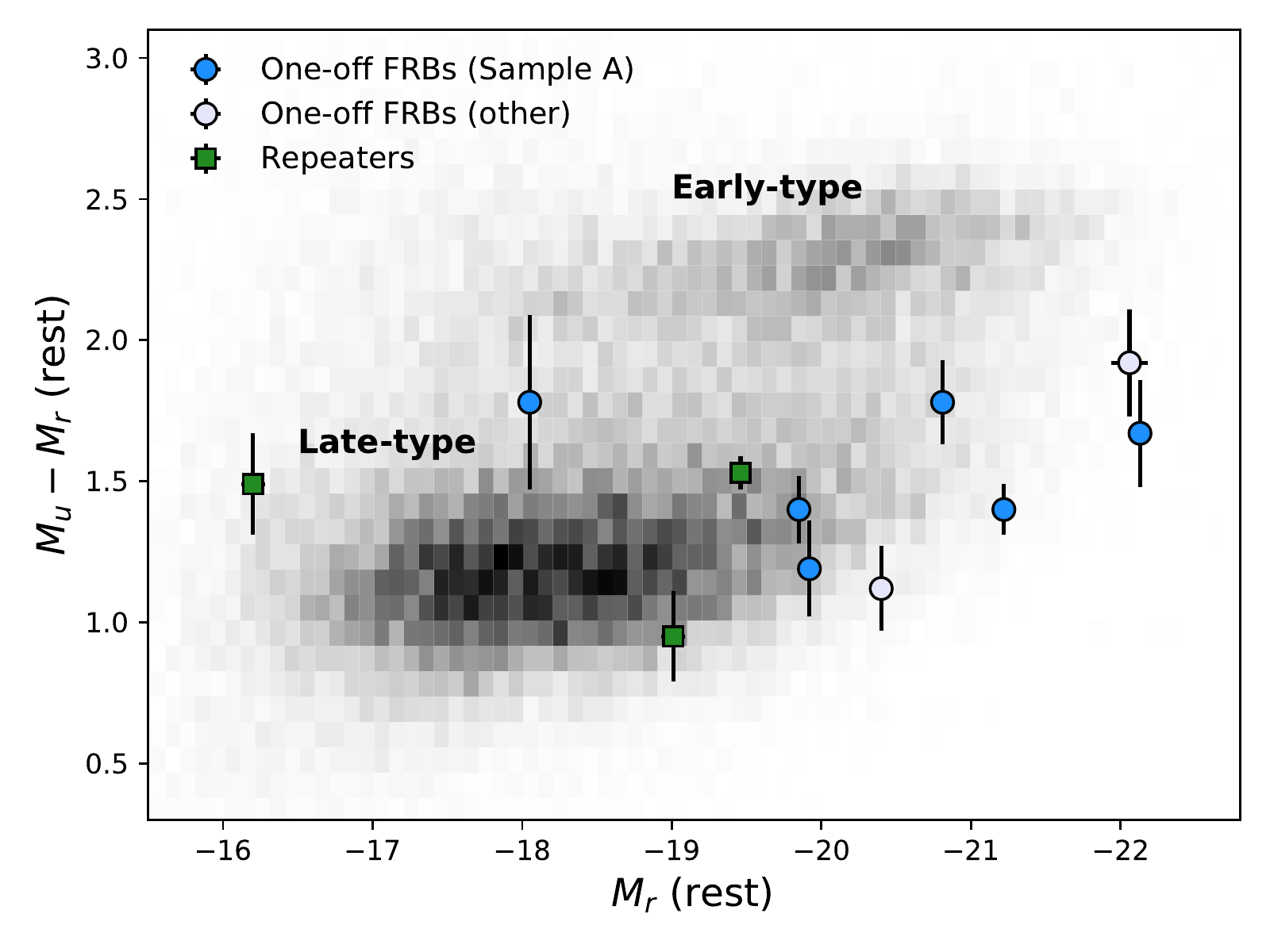}
    \caption{Rest-frame color-magnitude diagram of the host galaxies of repeating and nonrepeating FRBs compared to the underlying field galaxy population from the PRIMUS survey \citep{Moustakas13}. The FRB symbol notations are identical to Figure~\ref{fig:zlum}. The majority of the FRB hosts are part of the brightest galaxy population. 
    }
	\label{fig:colmag}
\end{figure}

We then consider the color-magnitude properties of the FRB hosts, which is a useful indicator of the overall stellar population in these galaxies. In Figure~\ref{fig:colmag} we compare the absolute $r$-band magnitudes $M_r$ and the rest-frame $M_u - M_r$ colors of the FRB hosts to the galaxies from the PRIMUS survey \citep{Moustakas13}, here representing the general population of $z < 0.5$ galaxies. We find that the majority of FRB-host galaxies sample the brighter region of the magnitude distribution, consistent with the initial sample studied in \citet{Bhandari20a}. This suggests that FRB hosts typically trace more massive galaxies than the underlying galaxy population (see also Section~\ref{ssec:mstarsfr}). Moreover, 
we note that the host galaxies of the three repeating FRBs are fainter than nearly all of the hosts of the apparently nonrepeating FRBs. 

Approximately half of the FRB-host galaxies have
colors consistent with the SF so-called ``blue cloud", similar to most late-type galaxies \citep{Strateva01}. 
The remainder (FRBs\,180916, 180924, 190523, 191001, and 200430) are located in the so-called ``green valley", the intermediate region between the two main populations. 
These galaxies may be transitioning from SF to the quiescent galaxies of the ``red sequence" \citep{Martin07}. 

Figure~\ref{fig:colmag} further reveals that the FRB hosts
do not populate either of the main loci of the blue or red
sequences. To quantify this impression,
we perform 2D Kolmogorov-Smirnov (KS) tests on the color-magnitude distribution of FRB-host galaxies (considering one-off, repeaters, and the full population) compared to the distribution of early- and late-type galaxies. 
The results are summarized in Table \ref{tab:colmag_KS}. 
The hypothesis that the full FRB population is drawn from the same underlying distribution as the full galaxy population is rejected with a KS probability $\mpks=0.007$. 
Other scenarios are rejected at higher significance levels ($\mpks <0.002$). 
Considering only the repeating FRBs, we find \pks\ values
consistent with those drawn from the late-type population.

\subsection{FRB Hosts in the BPT Diagram} \label{ssec:bpt}

\begin{figure}[!t]
\centering
    \includegraphics[width=8.7cm]{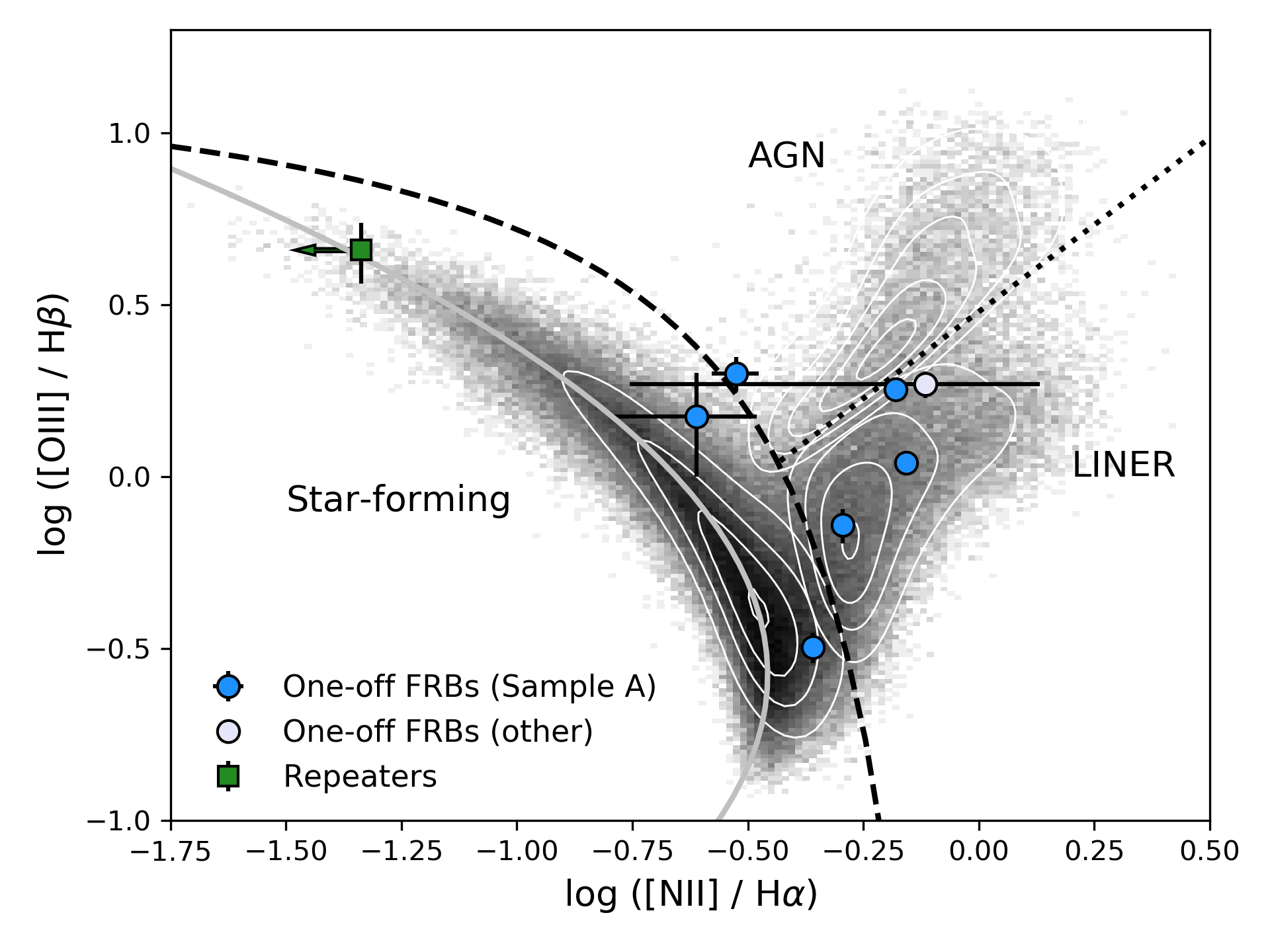}
    \caption{BPT \citep{Baldwin81} classification diagram for FRB hosts. The FRB symbol notations are identical to previous figures. The gray-scale background shows the density distribution of nearby ($0.02 < z < 0.4$) emission-line galaxies from the SDSS, only considering ${\rm S/N} > 5$. The solid gray line follows the highest density of local SF galaxies \citep{Brinchmann08}. The dashed and dotted black lines represent the demarcation line between SF galaxies and AGNs \citep{Kauffmann03} and AGN and LINERs \citep{CidFernandes10}, respectively. The white contours show the KDE of the galaxy population distributions used to model their individual PDFs.
    The majority of FRB hosts show excess line flux ratios compared to typical SF galaxies.
    }
	\label{fig:bpt}
\end{figure}

In Figure~\ref{fig:bpt} we show the [\oiii]/\hb\ and [\nii]/\ha\ nebular emission-line ratios of the FRB hosts in a Baldwin-Phillips-Terlevich (BPT) diagram \citep{Baldwin81}. This allows us to assess the dominant source of ionization and distinguish between typical SF galaxies, low-ionization nuclear emission-line region (LINER) galaxies, and AGNs \citep[see][for a recent review]{Kewley19}. 

\begin{table}[!t]
\begin{center}
\tabletypesize{\footnotesize}
\caption{$P$-values obtained via 2D KS tests with the null hypothesis that an FRB-host galaxy population (one-off, repeating, or all) is drawn from the same underlying distribution as early- or late-type galaxy populations or the full distribution.}\label{tab:colmag_KS}
\begin{tabular}{ l c c c } 
 \hline\hline
 Galaxy Type & $P_{\rm KS}$ (one-off) & $P_{\rm KS}$ (rep.) & $P_{\rm KS}$ (all) \\
 \hline
 All        & 0.002         & 0.178         & 0.007 \\
 Early-type & $<0.001$      & $<0.001$      & $<0.001$ \\ 
 Late-type  & $<0.001$      & 0.192         & $<0.001$ \\
 \hline
\end{tabular}
\end{center}
\end{table}

We have measured emission-line fluxes for the majority of the hosts in Sample~A, most of which were previously reported in \cite{Bhandari20a}. For comparison, we show the distribution of $\sim 75,000$ nearby ($0.02 < z < 0.4$) emission-line galaxies from the Sloan Digital Sky Survey (SDSS), with each emission line required to be detected at ${\rm S/N} > 5$. We also include the standard demarcation lines between SF, AGN, and LINER galaxies \citep{Kauffmann03,CidFernandes10}.

Examining Figure~\ref{fig:bpt}, we find that
the FRB hosts occupy a distinct region of the BPT diagram from
the dominant locus of SF galaxies:
the majority of FRB hosts show an excess in the [\nii]/\ha\ ratio compared to the ridge line tracing the highest density of local SF galaxies \citep{Brinchmann08}, with many 
located in the LINER region. The only exception is the host galaxy of the repeater FRB\,121102, which is located in the tail of the SF galaxy population. 

We use 2D KS tests to compare the FRB-host galaxy population (both with and without the repeater FRB\,121102) to each galaxy class. 
Galaxy classes are assigned according to the BPT diagram (Figure~\ref{fig:bpt}),
and the results are summarized in Table \ref{tab:BPT_KS}.
The FRB-host galaxy population is statistically 
inconsistent with the distribution of SF galaxies
($\mpks = 0.015$) and may favor the AGN+LINER
populations.

The excess of emission-line ratios from the locus of regular SF galaxies in the BPT diagram is generally attributed to a hard stellar ionizing radiation field or elevated ionization parameters of the interstellar medium \citep[ISM;][]{Brinchmann08,Steidel14}. The underlying emission mechanism is not completely clear, however. \citet{Thomas18} argue that the excess in line flux ratios can be described by an increased ``mixing" of AGN emission with the H\,{\sc ii} region emission. Alternatively, the same ionization effect can be produced by a dominating population of post-asymptotic giant branch (post-AGB) stars \citep{Yan12,Singh13}. 
The latter scenario aligns with the typical old stellar populations of the FRB hosts inferred from the SED modeling (see Section~\ref{ssec:sed} and Table~\ref{tab:hostprop}).
We do note, however, that the host galaxy of FRB\,190608 is found to contain a Type~I AGN based on the detection of broad \ha\ emission \citep{Stern12,Chittidi20}.
We do not detect similar broad \ha\ emission lines in the other FRB-host spectra. Ultimately, integral field unit (IFU) observations at high spatial resolution
of the host galaxies are needed to distinguish whether the central AGN or the overall LINER emission are the most common emission mechanisms producing the elevated ionization observed in FRB hosts.

\begin{table}[!t]
\tabletypesize{\footnotesize}
\caption{$P$-values obtained via a 2D KS test for FRB-host populations (considering the full set of repeating and nonrepeating bursts and the one-off bursts only) and different galaxy populations according to the BPT diagram.}\label{tab:BPT_KS}
\begin{tabular}{ l c c } 
 \hline\hline
 Galaxy Type & $P_{\rm KS}$ (one-off) & $P_{\rm KS}$ (all) \\
 \hline
 All            & 0.049         & 0.023 \\
 SF             & 0.004         & 0.015 \\ 
 AGN            & $< 0.001$     & $< 0.001$ \\
 LINER          & 0.041         & 0.012 \\
 AGN-LINER      & 0.122         & 0.044 \\
 SF-LINER       & 0.039         & 0.019 \\
 \hline
\end{tabular}
\end{table}


\subsection{Star-Formation Rates and Stellar Masses} \label{ssec:mstarsfr}

\begin{figure}[!t]
\centering
    \includegraphics[width=8.7cm]{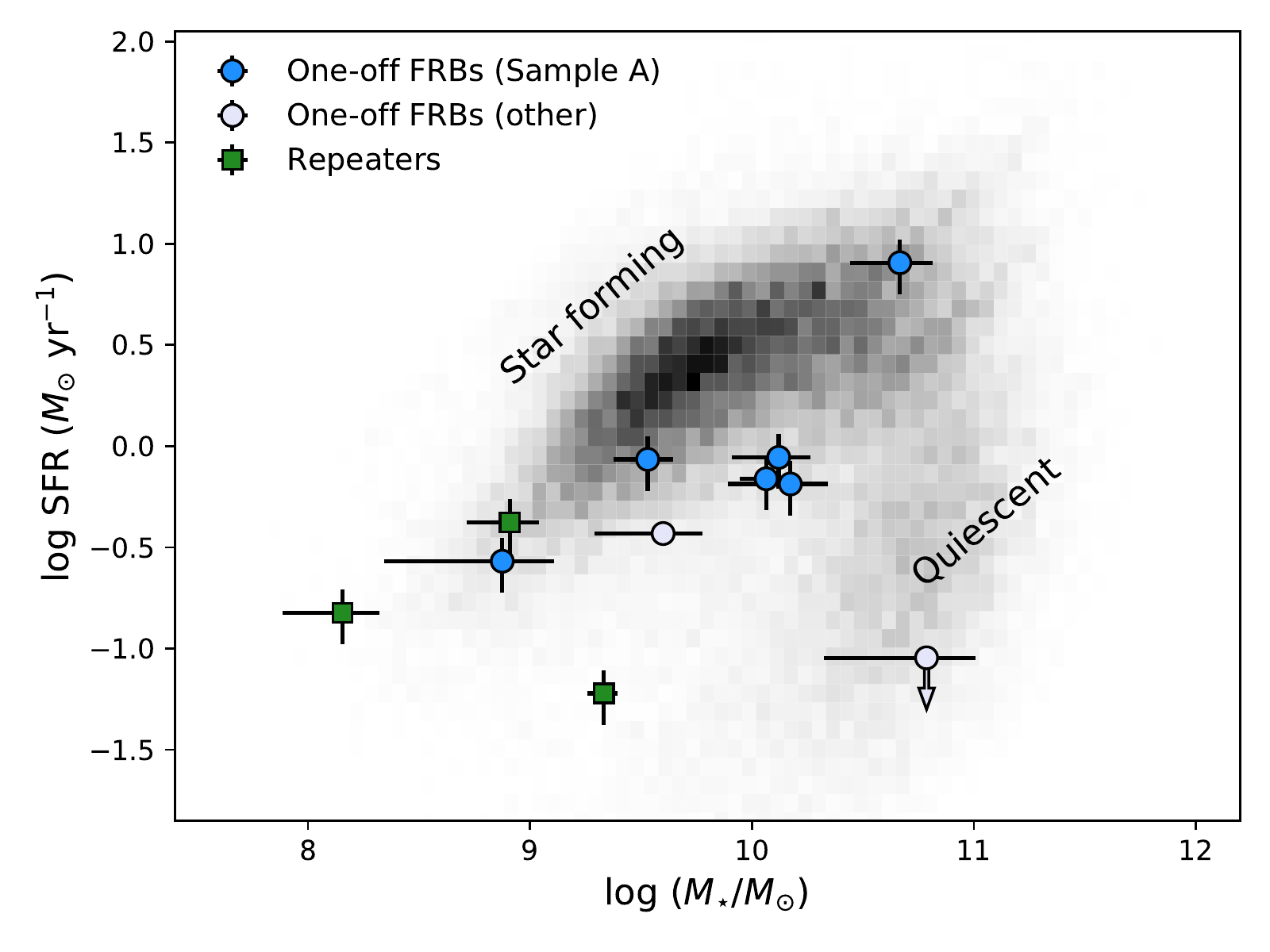}
    \caption{Star-formation rate vs. stellar mass $M_{\star}$ distribution of FRB hosts. The FRB symbol notations are again identical to previous figures, and we here also include the galaxies from the PRIMUS survey as the background sample. Because some of the hosts show LINER-like emission, the SFR should potentially be treated as an upper limit (see main text). The hosts of repeating FRBs show more diverse behavior: i.e., starbursts (FRB\,121102), regular SF (FRB\,190711) and quiescent (FRB\,180916) galaxies compared to the hosts of nonrepeating FRBs.
    }
	\label{fig:sfgms}
\end{figure}

We show the SFR$-M_{\star}$ distribution of the FRB hosts in Figure~\ref{fig:sfgms}. For the control sample, we again show the galaxies from the PRIMUS survey \citep{Moustakas13}. We caution that due to the LINER-like emission observed for most of the FRB-host galaxies, the SFR could in some cases only represent an upper limit on the actual rate because the total line emission might not solely reflect the star-formation activity.

Similar to the color-magnitude distribution 
(Figure~\ref{fig:colmag}; Section~\ref{ssec:colmag}),
we find that the FRB-host galaxies avoid the main sequence of SF galaxies (i.e.,\ the main locus of the control sample). Moreover, a 2D KS test yields a
low probability that the two distributions were
drawn from the same parent population ($\mpks <0.001$).
Intriguingly, the host galaxies of the known repeating FRBs show more diverse behavior than those hosting nonrepeating bursts, ranging from faint starburst (FRB\,121102), to regularly SF (FRB\,190711), and finally to quiescent (FRB\,180916) galaxies. The hosts of the repeating FRBs are all relatively low-mass galaxies ($M_{\star} < 2\times 10^{9}\,M_{\odot}$) compared to the overall FRB-host population (as described before in Section~\ref{ssec:colmag}). 

We now consider the hypothesis that FRBs track stellar
mass. Specifically, we compare the observed distribution
$f_{\rm FRB} (M_{\star})$ with the stellar mass function
of low-$z$ galaxies $\phi(M_\star)$ weighted by stellar mass,
i.e.\ $f_{\rm FRB} (M_{\star}) \propto M_\star \phi(M_\star)$.
For this analysis we assume the parameterization of $\phi(M_{\star})$ derived by \citet{Davidzon17} for galaxies at $0.2 < z < 0.5$ in the COSMOS field. 

\begin{figure}[!t]
\centering
    \includegraphics[width=8.7cm]{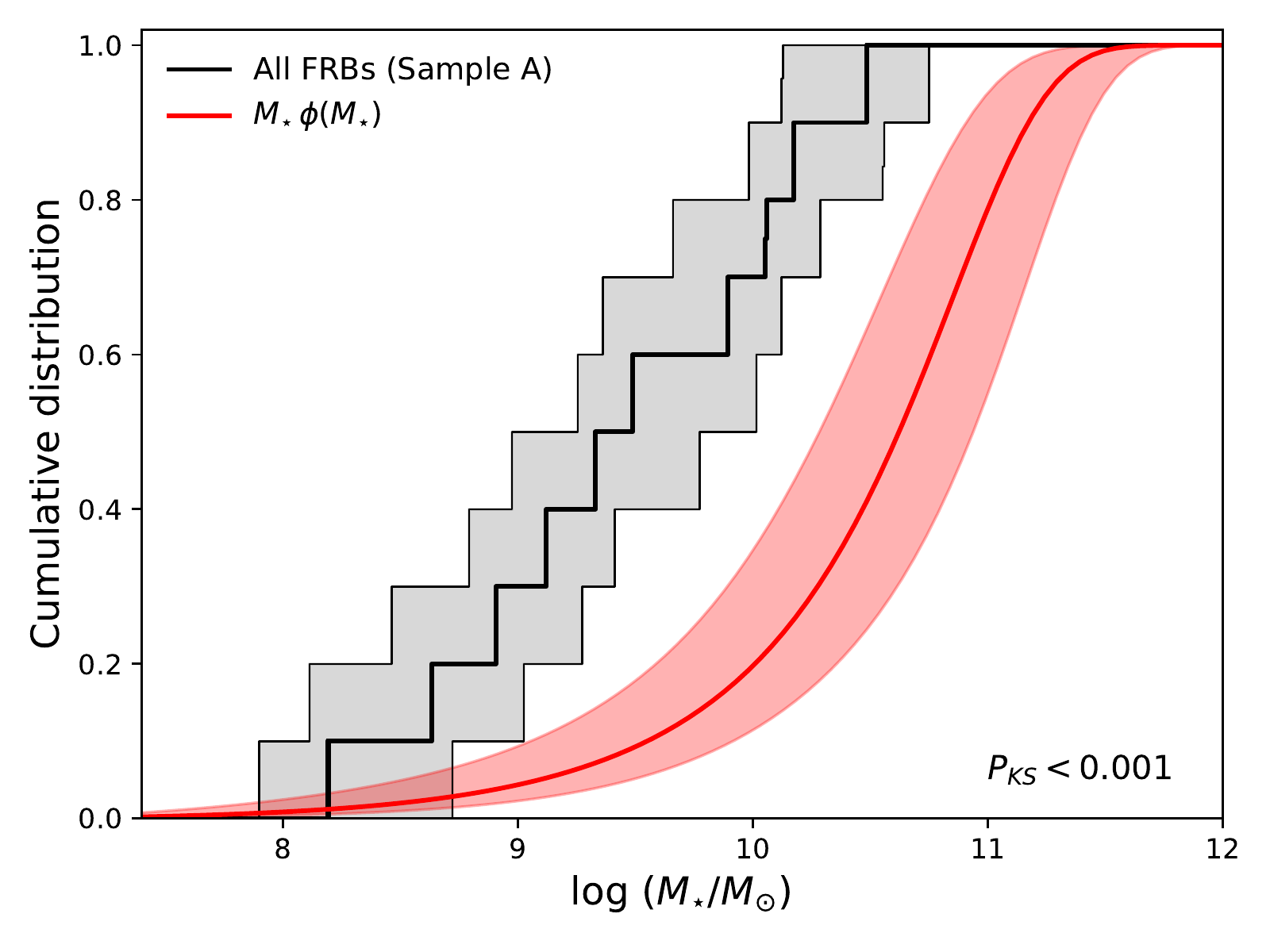}
    \includegraphics[width=8.7cm]{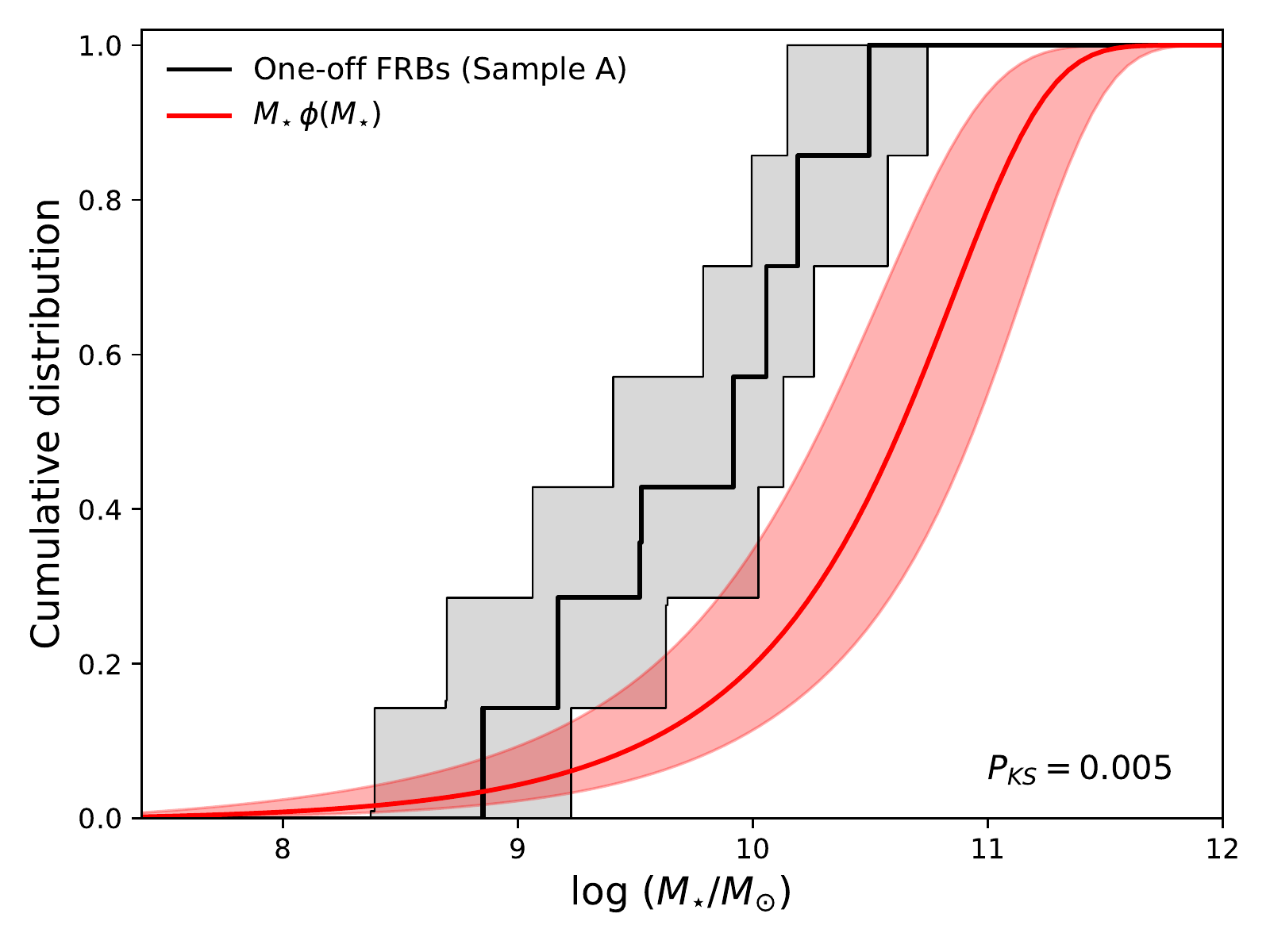}
    \caption{Stellar mass $M_{\star}$ cumulative distribution of all (top panel) and only the nonrepeating (bottom panel) FRB hosts in Sample A. The gray shaded region represents the $1\sigma$ uncertainty on the CDF, combining the error on the measurements and due to the sample size (see text for details). The observed distribution is compared to the stellar mass function $\phi(M_{\star})$ for the full COSMOS sample from \citet[][at $0.2 < z < 0.5$]{Davidzon17}, assuming a mass-weighted selection $M_\star \phi(M_{\star})$ from the field galaxy mass function (red line). 
    The computed $P$-values from a one-sided KS test between the distributions are listed at the bottom of both panels.
    }
	\label{fig:massweight}
\end{figure}

In Figure~\ref{fig:massweight} we plot the cumulative stellar mass distribution of the FRB hosts in Sample~A. We first consider all the hosts (top panel) and then only the hosts of the one-off FRBs (bottom panel). The uncertainty regions on the cumulative distribution functions (CDFs) are estimated by combining the two sources of uncertainty: the errors on the individual data points, and the error from the sample size. We calculate the former using Monte Carlo error propagation, assuming that the probability density function (PDF) of each data point is described by a Gaussian profile, with the standard deviation given by the error on the measurement \citep[similar to the procedure described in][]{Palmerio19}. We then estimate the median and $1\sigma$ confidence bounds on the CDF from 10,000 realizations of the data sampling. The error from the sample size is then computed via bootstrapping and added to show the combined uncertainty region.

A comparison of the CDF of all the FRB hosts 
to the mass-weighted stellar mass distribution of field galaxies $f_{\rm FRB} (M_{\star})$ yields a probability of $P_{\rm KS} < 0.001$ from a one-sided KS test for the two distributions to be drawn from the same underlying mass distribution.
Therefore these results rule out the null hypothesis that
FRBs directly track stellar mass. When we limit this to one-off
FRBs (Figure~\ref{fig:massweight}, lower panel) the offset is reduced, but the probability remains low.


\subsection{Mass-Metallicity Relation} \label{ssec:massmet}

In addition to the stellar mass and SFR, the gas-phase metallicity is a strong indicator of the present stellar populations and can thus also provide constraints on the most likely progenitor channels. Indeed, the typical low-metallicity environments of LGRB host galaxies were vital in the conception of the ``collapsar” progenitor model for LGRBs \citep[e.g.][]{Yoon06}. A more direct, quantitative comparison between FRB and LGRB hosts (in addition to the hosts of other types of transients) is provided in Sect.~\ref{ssec:comp}.

\begin{figure}[!t]
\centering
    \includegraphics[width=8.7cm]{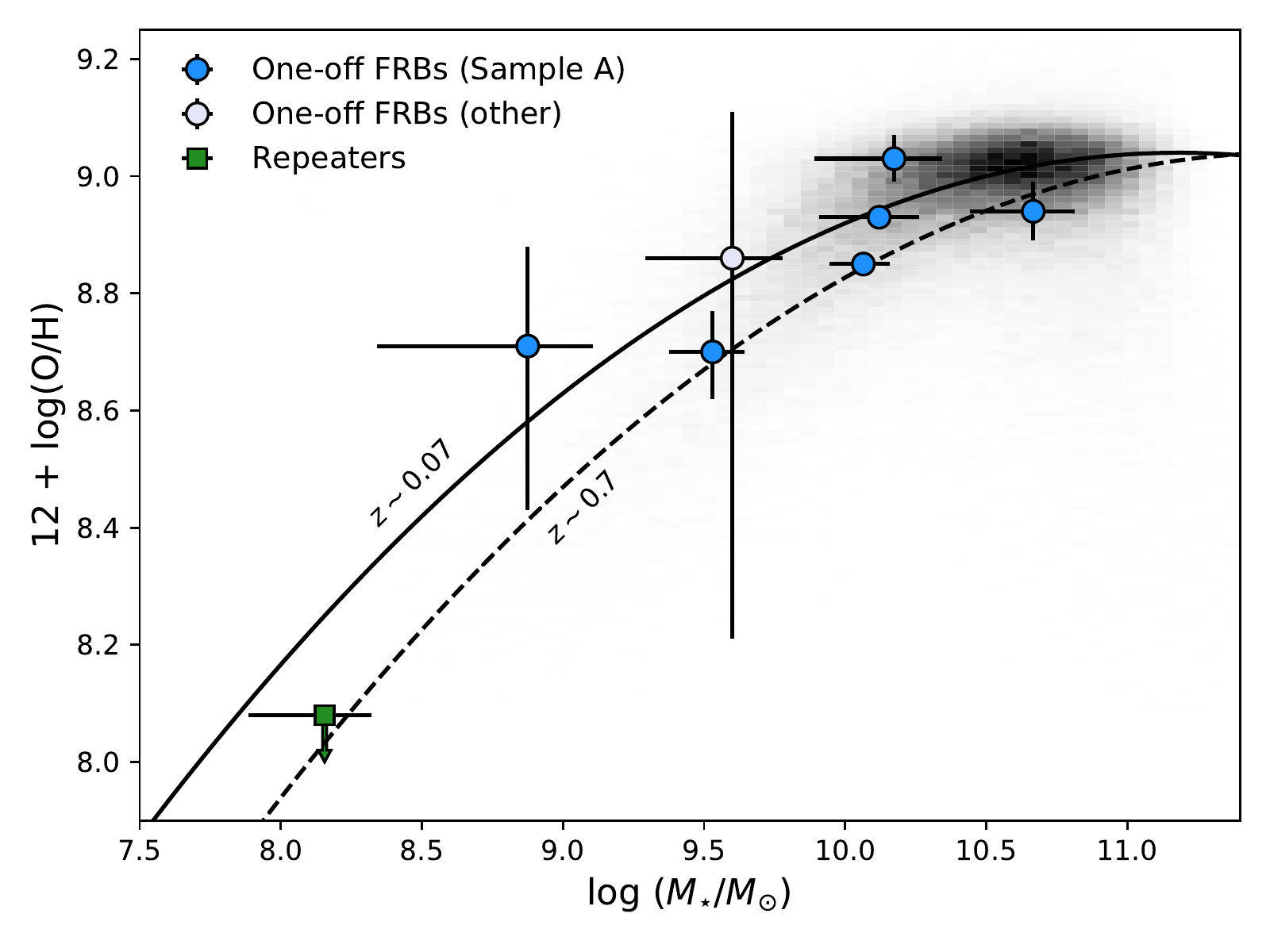}
    \caption{Mass-metallicity relation of FRB hosts. The FRB symbol notations are identical to previous figures. Here, we only show a subset of the FRB hosts for which sufficient emission-line fluxes have been measured to derive or place limits on the oxygen abundance $12+\log$(O/H), assuming the \citet{Hirschauer18} O3N2 calibration, see Table~\ref{tab:lineflux}. For the gray-scale background distribution, we again use the local SDSS emission-line sample. For reference, the mass-metallicity relations from \citet{Maiolino08} are shown for $z\sim 0.07$ (solid) and $z\sim 0.7$ (dashed). 
    }
	\label{fig:massmet}
\end{figure}

In Figure~\ref{fig:massmet} we show the metallicities of the FRB-host galaxies in terms of their oxygen abundances $12+\log$(O/H) as a function of stellar mass (i.e., the mass-metallicity relation). For the control sample, we show the SF galaxies from the SDSS emission-line sample, with metallicities calibrated using the same strong-line diagnostics as for the FRB hosts (see Section~\ref{ssec:met}). For comparison, we overplot the mass-metallicity relations at $z\sim 0.07$ and $z\sim 0.7$ by \citet{Maiolino08}. We find that the majority of FRB hosts is consistent with the $z\sim 0.07-0.7$ mass-metallicity relations and the underlying field galaxy population. 


\subsection{Locations: Projected Physical and Host-normalized Offsets}

Last we consider the projected physical offsets ($\rho$) of the expanded sample of FRBs, in addition to the projected offsets normalized by the half-light radii of the hosts ($\rho/R_{\rm eff}$).
When operating in the ICS mode, ASKAP/CRAFT can now deliver subarcsecond localizations of FRBs upon detection, without requiring the use of follow-up facilities on repeat bursts.
Both approaches allow us to accurately determine the 
FRB emission sites with respect to their host-galaxy centers (``offsets''), which provide additional clues to the progenitors of FRBs. Indeed, the offset distributions of other transients have provided a key diagnostic for understanding their origins (discussed further in Section~\ref{sec:frbprog}).

For each FRB in Sample A, we measure the angular offset between the FRB location and its host-galaxy center, taking into account positional and astrometric uncertainties for each measurement and use the redshift of the host galaxy to convert to physical offsets in kpc. We determine a broad range of projected physical offsets for the FRBs in Sample~A, spanning from 0.6\,kpc (FRB\,121102) to $\approx 11$\,kpc (FRBs\,190611 and 191001); they are listed in Table~\ref{tab:hostprop}. Overall, we find that FRBs have significant offsets relative to the centers of their host galaxies, with median and mean values of 3.3 and 4.8\,kpc, respectively. We caution that the observed FRB population presented here could be biased against small offsets due to an increasing effect of DM scattering or ``smearing" caused by the dense ISM, thus decreasing the FRB detection probability closer to their host-galaxy centers. However, we expect this effect to be minor. 
Using the derived host-galaxy sizes ($R_{\rm eff}$), we also measure the host-normalized offsets for Sample~A. 
We caution that most of the \Reff\ values were derived
from seeing-limited observations and are therefore subject
to significant uncertainty for the smaller galaxies
($\mReff \lesssim 1''$).
Nevertheless, we find a range of values, $\rho=0.4-5.3\,R_{\rm eff}$ with median and mean values of 1.4\,$R_{\rm eff}$ and 1.7\,$R_{\rm eff}$, respectively. 
We note that this is larger than the median expected offset if FRBs traced the locations of stars in their disks (e.g., $1\, \mReff$).


\section{Implications for FRB Progenitors} \label{sec:frbprog}

We have here shown that FRB hosts exhibit very diverse environments: in particular, we observe a large variety in terms of their morphologies, ranging from early- to late-type galaxies, and found that FRB hosts are characterized by a broad, continuous range of rest-frame colors, luminosities, stellar masses, SFRs and ages. We now explore the implications for the nature of FRB progenitors through further comparisons of their host-galaxy properties to the hosts of other astronomical transients.


\subsection{Comparisons to the Host Properties and Offset Distributions of Other Transients} \label{ssec:comp}

The progenitors of other known transients such as LGRBs, SGRBs, CC, and SNe Ia provide a natural baseline for comparison to the FRB-host population because these have been intensively studied, and have known or likely known progenitors. Investigating the connection between their hosts and galaxies hosting FRBs can therefore provide important (though indirect) clues to the most likely FRB progenitor channels. Based on the first small samples of FRB-associated hosts \citep{Bhandari20a,Li20}, it was already evident that the majority had generally high masses and low SFRs (excluding FRB\,121102). Our work has further cemented this picture based on a sample of \nsampA\ secure host galaxies.

\subsubsection{Luminosity, SFR, and Stellar Mass}

Here, we further discuss the connection between FRB hosts and those of other astronomical transients and compare them quantitatively. The typically high luminosities and stellar masses but modest SFRs observed in this work are generally consistent with the galaxy populations hosting CC-SNe and SGRBs, which are found to predominantly occur in luminous, massive galaxies \citep{Prieto08,Berger09,Kelly12,Taggart19}. These physical properties are in stark contrast to the typically elevated specific SFRs (sSFR = SFR/$M_{\star}$) observed for the hosts of LGRBs. Moreover, the host galaxies of LGRBs at $z \lesssim 1$ are typically at the faint, low-mass end of the SF galaxy population \citep{Savaglio09,Vergani15,Schulze15,Perley16}, in contrast to what we have observed for the majority of FRB hosts.

\begin{figure}[!t]
\centering
    \includegraphics[width=8.7cm]{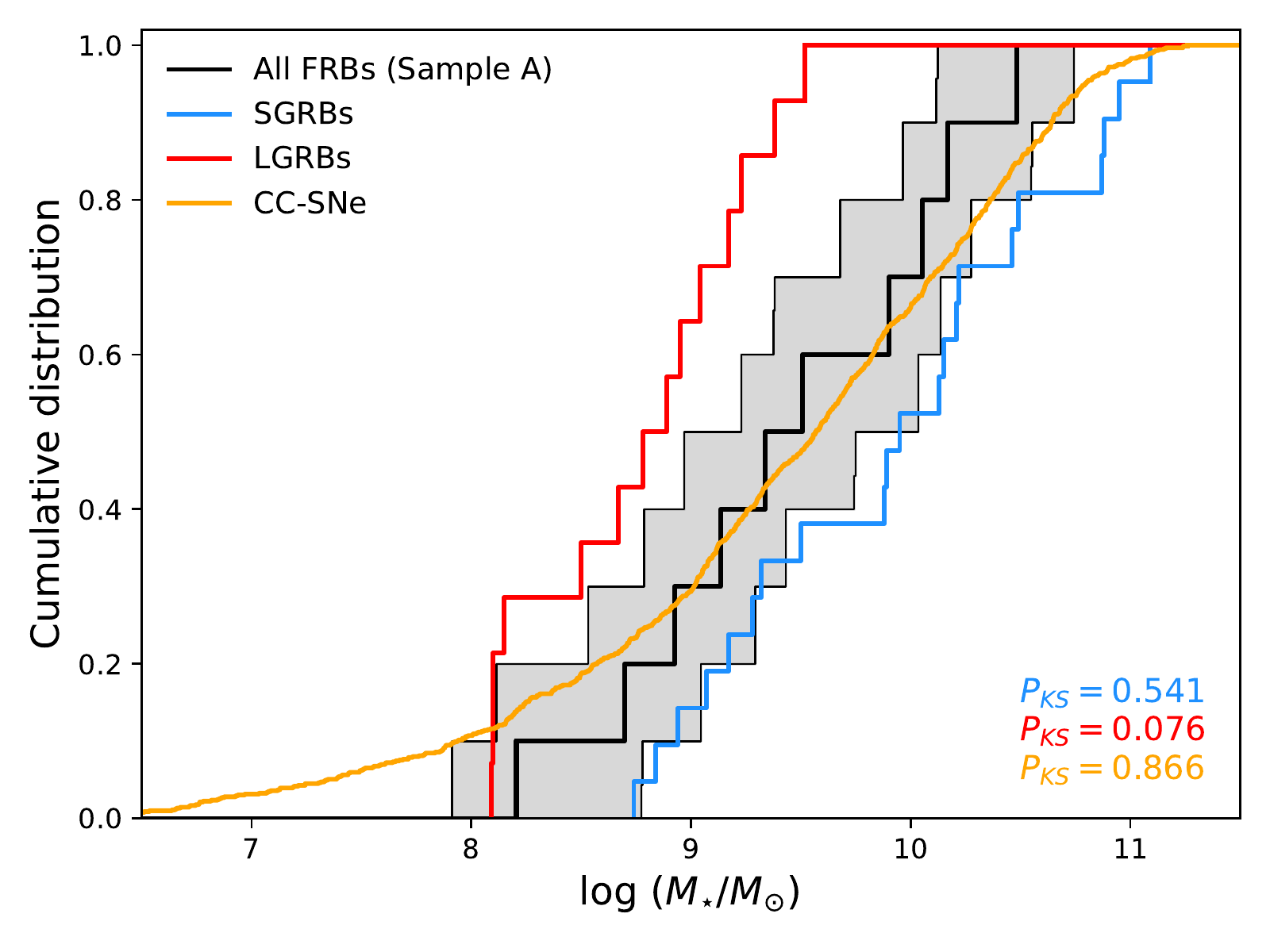}
    \includegraphics[width=8.7cm]{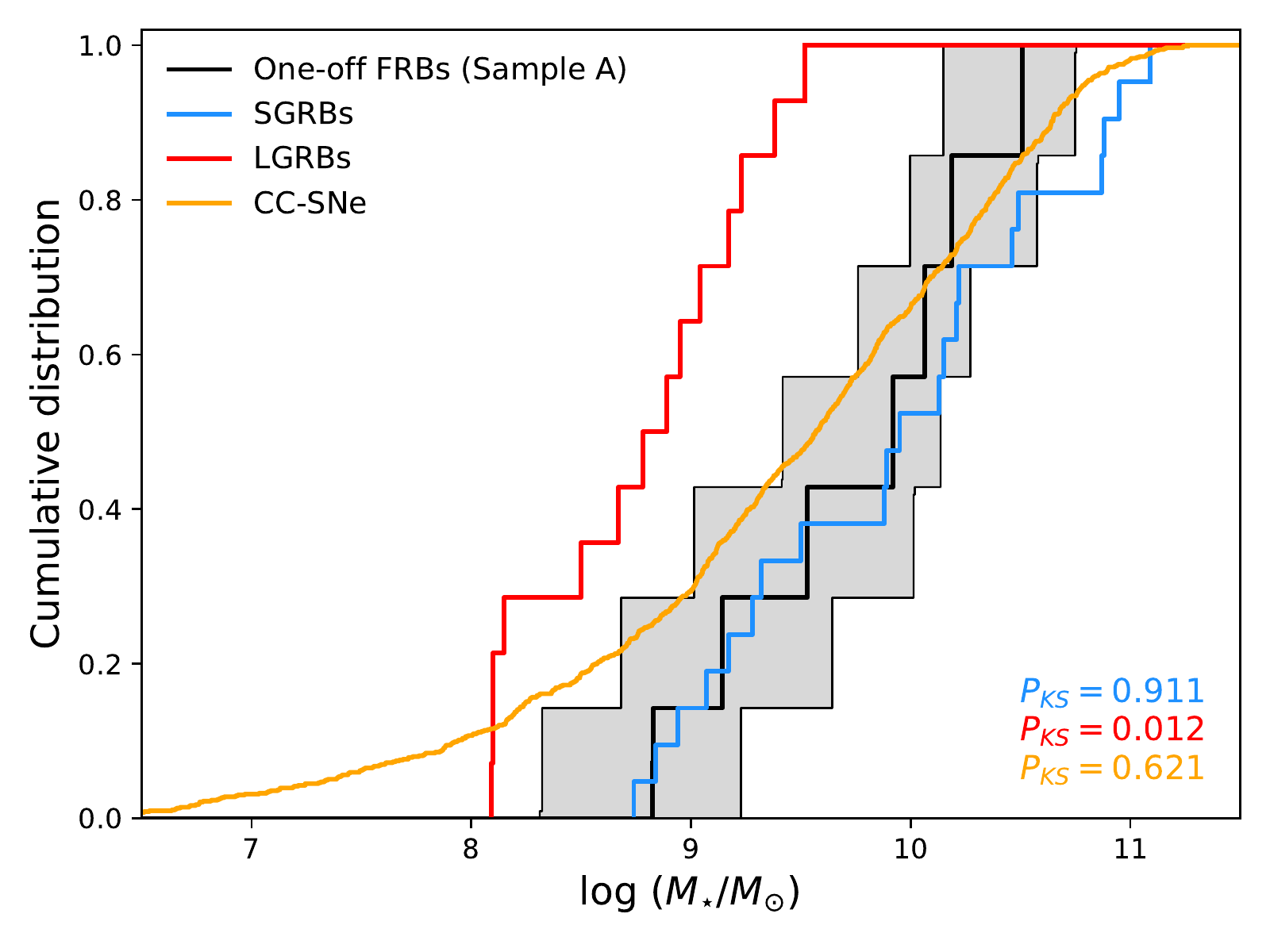}
    \caption{Stellar mass $M_{\star}$ cumulative distribution of all (top panel) and only the nonrepeating (bottom panel) FRB hosts in Sample A. The gray shaded region represents the $1\sigma$ uncertainty on the CDF, combining the error on the measurements and that due to the sample size (see Section~\ref{ssec:mstarsfr} for details). For comparison, we overplot samples of SGRB \citep[blue;][]{Nugent20}, LGRB \citep[red;][]{Vergani15}, and CC-SN \citep[orange;][]{Schulze20} hosts, all at $z<1$. The computed $P$-values from a two-sided KS test are listed for the comparison distribution functions relative to the FRB samples.
    }
	\label{fig:massdistcomp}
\end{figure}

In Figure~\ref{fig:massdistcomp} we compare the stellar mass distribution of the FRB hosts to the host galaxies of these other transients, namely SGRBs \citep{Nugent20}, LGRBs \citep{Vergani15}, and CC-SNe \citep{Schulze20}. To mitigate the effects of the cosmological evolution of galaxies and ensure a fair comparison, we require $z < 1$ for the host galaxies for all of the comparison samples.
The \mstar\ CDF for FRBs is intermediate between the 
lower mass hosts of LGRBs and the higher mass hosts
of SGRBs, and most closely following those of CC-SNe. Neither of the SGRB or CC-SNe host populations, however, has statistically inconsistent CDFs. When the comparison is restricted to the one-off FRBs, the correspondence to the host galaxies of LGRBs is further disfavored with
$\mpks < 0.05$.

\subsubsection{No Evidence for Metal Aversion in FRB Hosts}

As was demonstrated in Section~\ref{ssec:massmet}, the stellar masses and metallicities of FRB hosts are generally consistent with the mass-metallicity relations observed for field galaxies at $z = 0.07 - 0.7$. This is again consistent with the host galaxies of SGRBs at $z<1$ \citep{Berger09} and CC (Type II)/SNe Ia \citep{Prieto08}, which are also found to closely track the mass-metallicity or luminosity-metallicity relations of field galaxies at similar redshifts. In contrast, the production of LGRBs appears to be heavily suppressed in more metal-rich environments \citep[at least at $z<1$;][]{Perley16} compared to field galaxies at similar masses. 
FRB progenitors show no such metallicity bias in their host galaxies.

\begin{figure*}[!t]
\centering
    \includegraphics[width=8.7cm]{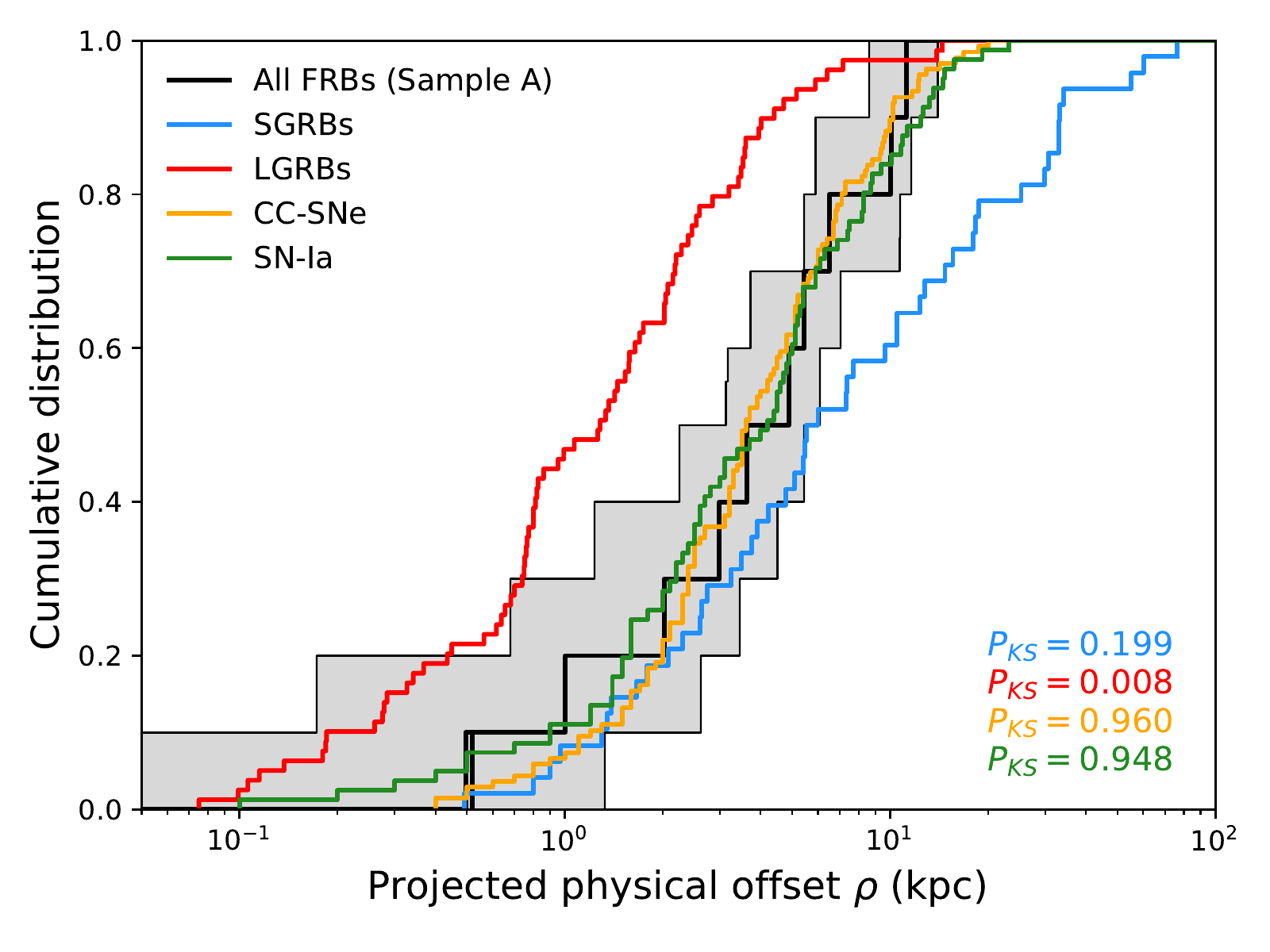}
    \includegraphics[width=8.7cm]{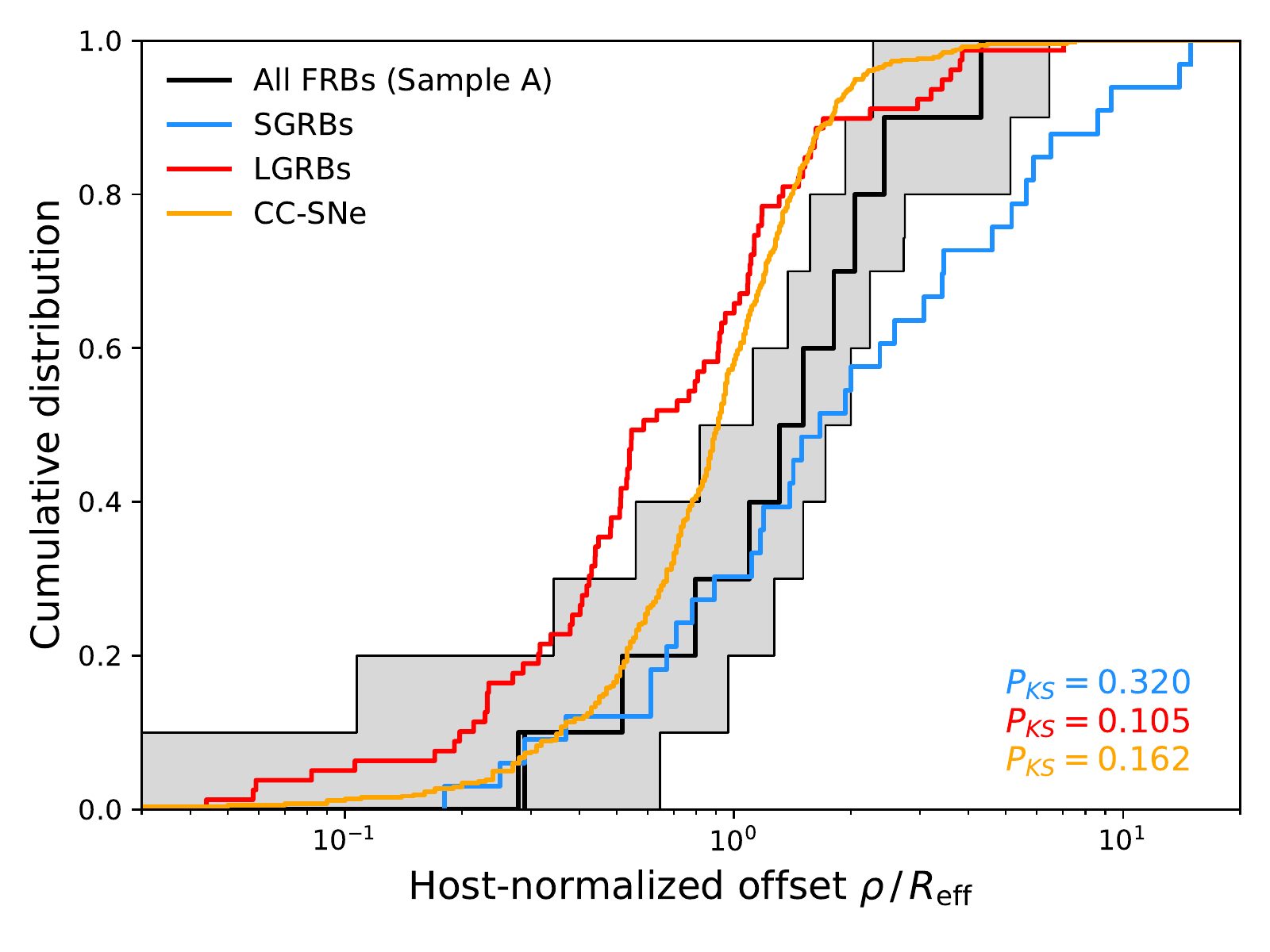}
    \caption{Cumulative distribution of projected physical offsets $\rho$ (left) and host-normalized offsets $\rho/R_{\rm eff}$ (right) for all FRBs (in Sample~A). The gray shaded region again represents $1\sigma$ uncertainty on the CDF, combining the error on the measurements and that due to the sample size. Reference samples of SGRBs \citep[blue;][in preparation]{Fong10,Fong13,Fongprep}, LGRBs \citep[red;][]{Blanchard16}, CC-SNe \citep[orange;][]{Prieto08,Kelly12}, and SNe Ia \citep[green;][]{Prieto08} are shown for comparison. The effective radii of the SNe Ia were not available to determine their host-normalized offset distribution. The computed $P$-values from a two-sided KS test are again listed for the comparison distribution functions relative to the total FRB sample. The FRB projected physical offset distribution is intermediate between the LGRB and SGRB distributions, but closely follows the CC and SN Ia offset distributions. When only the host-normalized offsets are considered, the distributions between the different types of transients becomes less distinct.
    }
	\label{fig:offs}
\end{figure*}

\subsubsection{Physical and Host-normalized Offsets}

We then compare the physical and host-normalized offset distributions of FRBs to that of the SGRB \citep[][in preparation]{Fong13,Fongprep}, LGRB \citep{Blanchard16}, and CC/SNe Ia \citep{Prieto08,Kelly12} populations in Figure~\ref{fig:offs}. For context, the locations of LGRBs match expectations for the H\,{\sc ii} regions of massive stars in an exponential disk, commensurate with their massive star progenitors \citep{Bloom02,Blanchard16,Lyman17}. For SGRBs, which originate from older stellar populations, the broad range of projected offsets of SGRBs extending to tens of kiloparsec are believed to result from neutron star kicks and delay times, providing a strong link to their neutron star merger progenitors \citep{Fong13}. 

In comparison, it is already evident that most FRBs are not coincident with the nucleus of their hosts, disfavoring models involving AGN or supermassive black holes in general \citep[see Figure~\ref{fig:images}; and also][]{Bhandari20a}. The offset distribution of the full sample of FRBs is found to closely follow the observed distribution of CC and SNe Ia, with one-sided KS tests yielding $P_{\rm KS} = 0.96$ and $P_{\rm KS} = 0.95$, respectively. The FRB and SGRB offset distributions are also consistent, but a larger fraction of SGRBs are observed to occur at even greater distances from their host-galaxy centers \citep{Fong13}, which is more consistent with theoretical expectations of binary neutron star mergers \citep{FryerKalogera97,Bloom99,Belczynski06}. 
We find similar results when we restrict our analysis to the hosts of
one-off FRBs. On the other hand, LGRBs are observed to be one of the most centrally concentrated populations, which is inconsistent with that of FRB hosts (with $P_{\rm KS} = 0.008$).

When the offsets are normalized by the effective radii of the host galaxies (Figure~\ref{fig:offs}, right), the results are qualitatively
similar, although the LGRB distribution is no longer 
inconsistent at high confidence.

\subsubsection{Implications for FRB Progenitor Channels}

Because the progenitors of FRBs (and astronomical transients in general) are linked to specific stellar populations and galaxy environments, determining these host-galaxy properties also allows us to place constraints on the likely progenitor channels of FRBs. Based on the similar properties of the first identified FRB host (that of the repeater, FRB\,121102) to those of LGRBs and SLSNe \citep{Tendulkar17}, a possible common progenitor channel of ``young" magnetar remnants producing FRBs was proposed \citep{Metzger17,Nicholl17}. FRB\,121102 originated from a low-mass and intensely SF galaxy relative to the typical FRB-host-galaxy population \citep{Li19,Bhandari20a}. With the addition of two repeaters, FRBs\,180916 and 190711, as well as the expanded sample of the hosts of apparent one-off FRBs studied here, the host properties exhibit a more continuous function in terms of their luminosities, stellar masses, and SFRs than previous studies. In the context of the sample of 13 hosts studied here, we note that FRB\,121102 is still on the extreme end in terms of its host properties (e.g., metal-deficient, low stellar mass, high SFR, and low luminosity). 

Finally, a metal deficit and high SFR per unit stellar mass seem to be crucial for the probability of a nearby ($z<1$) galaxy to host LGRBs or SLSNe. No such restrictions appear to govern the production of FRBs. We can thus conclude that the rapidly rotating massive stars that are believed to produce LGRBs \citep{Woosley06} are unlikely to constitute the majority of progenitors producing FRBs \citep{Marnoch20}. This conclusion is supported by the presence of some FRBs in galaxies with older stellar populations.
 
Instead, out of the other progenitor models proposed thus far, the physical properties of the FRB hosts are most similar to the host populations of SGRBs and CC/SNe Ia \citep{Bhandari20a,Li20}, although with a larger fraction of quiescent, older galaxies than the CC-SNe host population. Indeed, FRBs have been proposed to originate from either young or long-lived stable magnetars produced by a variety of channels, including binary neutron star mergers (at least some of which produce SGRBs) or in the accretion-induced collapse (AIC) of white dwarfs \citep{Moriya2016,Margalit19}. The FRB-host properties presented here, which span the full range of properties occupied by field galaxies at similar redshifts, are currently consistent with a single progenitor that can accommodate a diverse set of host properties, although contributions from multiple progenitors cannot be ruled out. Coupled with the similarities found when compared to both SGRB and SNe Ia host properties, magnetars produced as a result of binary neutron star mergers and/or white dwarf scenarios (e.g., \citealt{Kashiyama2013,Moriya2016}) remain viable FRB progenitor channels. In conclusion, it is clear that there is not a single preferred type of host galaxy (as is the case for LGRBs or SLSNe), and that any progenitor model will have to confront the diverse set of galaxy properties exhibited by the FRB-host population, as well as the relationships to the underlying galaxy population.


\subsection{Is the Milky Way a Typical FRB Host?}

An intriguing new clue to the origin of FRBs came from the discovery of a brief radio burst from the Galactic magnetar/soft gamma repeater SGR\,1935+2154 \citep{Scholz20,Bochenek20a}. The released radio energy in this event is approximately two orders of magnitude lower than that observed for the weakest of the cosmological FRBs \citep{CHIMEGalFRB,Bochenek20b}, and approximately five orders of magnitude fainter than typical one-off bursts (such as FRBs\,180924 and 181112). This is not unexpected, however, because such weak bursts would be difficult to detect at $z\gtrsim 0.1$, and this Galactic ``FRB" might therefore represent the faint end of the observed extragalactic FRB distribution.

To further examine this possibility, we can now pose the question whether the Milky Way is a typical FRB-host galaxy. Based on the inferred properties of $M_{\rm \star,MW} = (6.08\pm 1.14)\times 10^{10}\,M_{\odot}$ and SFR$_{\rm MW} = 1.65\pm 0.19\,M_{\odot}$\,yr$^{-1}$ \citep{Licquia15}, we find that the Milky Way is indeed consistent with being a typical FRB-host galaxy. Its SFR$-M_{\star}$ relation places it in the transition region between SF and quiescent galaxies, intermediate to what has been observed for the most massive ($M_\star > 10^{10}\,M_{\odot}$) FRB hosts. The same is true for the mass-metallicity relation of the Milky Way. Our Galaxy is thus a typical FRB-host galaxy. In addition, SGR\,1935+2154 is located approximately 9\,kpc from the Galactic center \citep{Kothes18,Bochenek20b}, which is in the high end of the physical offset distribution of the burst sites observed here. All these considerations further support the connection between the radio emission from SGR\,1935+2154 and low-luminosity FRBs. 

\section{Summary and Outlook} \label{sec:conc}

Here, we have presented new observations of five host galaxies of arcsecond-localized FRBs, together with a comprehensive analysis of their stellar population properties (colors, metalliticies, luminosities, stellar masses, mass-weighted ages, and SFRs) and locations with respect to their host-galaxy centers. One of these FRBs (FRB\,190711) is confirmed to be a repeating event, whereas the remaining four are apparently nonrepeating. 

We explored these properties in the context of all previously reported hosts of well-localized events: 10 FRB-host galaxies detected by CRAFT/ASKAP, as well as 3 additional FRBs discovered by other surveys (CHIME, DSA, and Arecibo). The precise localizations of these events enabled us to determine the most likely host-galaxy associations and define a “gold” sample (Sample A) of 10 hosts on which we based our statistical analyses. Of these hosts, 3 are known repeaters and 7 are (apparently) one-off bursts. To homogenize the results and present a uniform analysis that is not subject to systematic differences in stellar population modeling across several works, we obtained additional spectroscopic and photometric data of all 10 FRB hosts in Sample A and presented new SED modeling and spectral fits based on these observations
(including the 4 host galaxies presented in \citealt{Bhandari20a}). 
This work represents the largest sample of FRB-host galaxies to date and enables a statistical examination relative to the field galaxy population and to hosts of other transient types, as well as an exploration between known repeaters and (apparently) one-off bursts. The main results were as follows.

\begin{itemize}
    \item The majority of FRB hosts populate the range between $L \sim 0.1L^* - L^*$ at $z < 0.7$. 
    We find a tendency for the hosts of known repeating FRBs to exhibit colors of late-type galaxies, and to be overall less luminous and have lower stellar masses than the average FRB host.
    \item The full sample spans a large, continuous range in color ($M_u-M_r = 0.9 - 2.0$), mass-weighted stellar population age ($0.06 - 1.6$\,Gyr), stellar mass ($M_\star = 10^{8} - 6\times 10^{10}\,M_{\odot}$), and
    SFR (${\rm SFR} = 0.05 - 10\,M_{\odot}\,{\rm yr}^{-1}$) spanning the full parameter space occupied by $z<0.5$ galaxies. While the hosts of nonrepeating FRBs are typically more massive than the average population of galaxies on the SF main sequence, the hosts of repeating FRBs exhibit much more diversity, ranging from starburst (FRB\,121102), to regular SF (FRB\,190711), to more quiescent (FRB\,180916) galaxies.
    \item Statistical tests revealed that the mass distributions of the FRB hosts can be ruled out 
    ($>99.9\%$\,c.l.) as being uniformly drawn from the underlying mass distribution of field galaxies at a similar redshift range. This implies that FRBs do not directly track stellar mass.
    \item The majority of FRB hosts are emission-line galaxies, but with line ratios in the BPT diagram that do not track the distribution of regular field galaxies. In particular, the FRBs exhibit a high
    incidence of cases in the LINER population, which is
    indicative of a harder radiation field. 
    \item The overall sample of FRB hosts were found to be metal rich, with oxygen abundances distributed between $12+\log{\rm (O/H)} = 8.7 - 9.0$ (the exception being the host of FRB\,121102 with $12+\log{\rm (O/H)} < 8.08$), although all are consistent with the $z\sim 0.07-0.7$ mass-metallicity relations and the underlying field galaxy population.
    \item The physical offsets from the FRB position to the host-galaxy centers range from 0.6 to 11\,kpc, with a median value of 3.3\,kpc. Normalizing these by the half-light radii \Reff\ of the host galaxies yields host-normalized offsets ranging from 0.4 to $5.3\,R_{\rm eff}$ with a median of 1.4\,$R_{\rm eff}$.
    \item Comparing the host-galaxy properties and the projected physical offsets of FRBs to those observed for the populations of other transients (SGRBs, LGRBs, and CC, and SNe Ia) allowed us to place empirical constraints on potential stellar transients accompanying (or preceding) the FRB events. We found that the generally massive and metal-rich environments of the FRB hosts disfavor similar progenitor channels to those producing LGRBs (with FRB\,121102 being the exception). Moreover, the FRB host-burst offset distribution is consistent with those observed for SGRBs, CC, and SNe Ia, and further disfavors LGRBs ($>99\%$\,c.l.).
    \item Based on the host properties alone, magnetars formed via binary neutron star mergers, accretion-induced collapses of white dwarfs, or regular CC-SNe are thus amongst the current most plausible mechanisms for the majority of the FRB population. Any progenitor models also have to accommodate the broad, continuous range of host-galaxy properties, likely reflecting a large variety in type or lifetime of FRB progenitors.
\end{itemize}

This work highlights the crucial role of $\sim$arcsecond-level localizations in making robust associations with host galaxies \citep[e.g.,][]{Eftekhari17} and delineating the progenitors of FRBs. The associations are made even more challenging by the fact that the FRBs localized thus far exhibit substantial offsets from their host galaxies. It is particularly notable that one-off FRBs do not appear to be drawn from more typical galaxies. If so, a prior may be to search less ordinary galaxies to determine the hosts of FRBs that are poorly localized. In the future, larger samples of FRB-host galaxies will further establish whether the typical physical properties discovered here are common for the full population of FRBs. In particular, a larger number of galaxies hosting known repeating FRBs needs to be observed to decisively conclude the relationship between known repeating and apparently nonrepeating FRBs, as gleaned from their host-galaxy environments. Because FRBs are observed to originate in a diversity of galaxies, it is also crucial to study the local environments of their burst sites. Understanding the FRBs and their environments better will also significantly aid in their use as cosmological probes.

\section*{Acknowledgements}
We would like to thank the referee for a thorough and constructive report, greatly improving the presentation of the results from this work.
We would also like to thank Johan P. U. Fynbo and Enrico Ramirez-Ruiz for enlightening discussions on the implications for FRB progenitor models and Jesse Palmerio for his insight into statistical modeling. We would also like to thank Vikram Ravi for sharing his images of the host galaxy and localization region of FRB\,190523.
K.E.H. acknowledges support by a Project Grant (162948--051) from The Icelandic Research Fund. The Fast and Fortunate for FRB
Follow-up team acknowledges support from 
NSF grants AST-1911140 and AST-1910471. N.T. acknowledges support by FONDECYT grant 11191217. W.F. acknowledges support by the National Science Foundation under grant No. AST-1814782. K.A. acknowledges support from NSF grant AAG-1714897. A.T.D. is the recipient of an ARC Future Fellowship (FT150100415). R.M.S. acknowledges support through ARC Future Fellowship FT190100155 and discovery project DP180100857. 
This work is partly based on observations collected at the European
Southern Observatory under ESO programs 0103.A-0101(A) and 0103.A-0101(B).
This work is partly based on observations obtained at the international Gemini Observatory, a program of NSF’s OIR Lab, which is managed by the Association of Universities for Research in Astronomy (AURA) under a cooperative agreement with the National Science Foundation, on behalf of the Gemini Observatory partnership: the National Science Foundation (United States), National Research Council (Canada), Agencia Nacional de Investigaci\'{o}n y Desarrollo (Chile), Ministerio de Ciencia, Tecnolog\'{i}a e Innovaci\'{o}n (Argentina), Minist\'{e}rio da Ci\^{e}ncia, Tecnologia, Inova\c{c}\~{o}es e Comunica\c{c}\~{o}es (Brazil),
and Korea Astronomy and Space Science Institute (Republic of Korea). The Gemini data were obtained from program GS-2019B-Q-132, and processed using the Gemini {\sc Pyraf} package\footnote{\url{https://www.gemini.edu/sciops/data-and-results/processing-software}}.
The Australian Square Kilometre Array Pathfinder and Australia Telescope Compact Array are part of the Australia Telescope National Facility, which is managed by CSIRO. Operation of ASKAP is funded by the Australian Government with support from the National Collaborative Research Infrastructure Strategy. ASKAP uses the resources of the Pawsey Supercomputing Centre. Establishment of ASKAP, the Murchison Radio-astronomy Observatory, and the Pawsey Supercomputing Centre are initiatives of the Australian Government, with support from the Government of Western Australia and the Science and Industry Endowment Fund. We acknowledge the Wajarri Yamatji as the traditional owners of the Murchison Radio-astronomy Observatory site. Partly based on observations made with the Nordic Optical Telescope, operated by the Nordic Optical Telescope Scientific Association at the Observatorio del Roque de los Muchachos, La Palma, Spain, of the Instituto de Astrofisica de Canarias.
Some of the data presented herein were obtained at the W. M. Keck Observatory, which is operated as a scientific partnership of the California Institute of Technology, the University of California and the National Aeronautics and Space Administration. The Observatory was made possible by the generous financial support of the W. M. Keck Foundation. The authors wish to recognize and acknowledge the very significant cultural role and reverence that the summit of Maunakea has always had within the indigenous Hawaiian community.  We are most fortunate to have the opportunity to conduct observations from this mountain. W. M. Keck Observatory access for FRB200430 was supported by Northwestern University and the Center for Interdisciplinary Exploration and Research in Astrophysics (CIERA).


\bibliography{ref}
\bibliographystyle{aasjournal}

\appendix

\section{Updated Literature FRB Host Properties}
\label{sec:literature}

\subsection{ASKAP/CRAFT}

\noindent 
{\it FRB\,180924, FRB\,181112, FRB\,190102, FRB\,190608, and FRB\,191001:}
All of these FRBs and their host galaxies were presented
in previous CRAFT publications \citep{Bannister19,Prochaska19b,Macquart20,Chittidi20,Bhandari20a,Bhandari20b}.
Several of the FRB coordinates have been improved
from refined analysis of the saved baseband data, enabling more precise positions for the FRBs and/or a better estimate of the astrometric registration of the FRB image \citep{Day20}. For FRB\,190102 we also include the {\it HST}/F160W band photometry measured by \citet[][{\it in prep.}]{Manningsprep} of $m_{\rm F160W} = 20.45\pm 0.01$\,mag in the SED fit of the host galaxy to improve the estimates of the stellar population parameters. We note that for FRB\,191001, the R.A. uncertainty reported in \cite{Bhandari20b} of 0.006s was incorrectly calculated, and should have been 0.02s (the latter adopted in this work).
All of these except for FRB\,181112 are included in Sample~A.
While we still maintain the association of FRB~181112
to DES\,J214923.66$-$525815.28 to be highly secure, the
foreground galaxy studied in \cite{Prochaska19b}
is sufficiently bright to also give
$\mpchance < 0.05$.  
Therefore this association places it within Sample~C.
A future Bayesian framework for FRB-host associations will
enable a direct comparison of the probabilities of the two
sources, and we expect that DES\,J214923.66$-$525815.28 
will by highly favored.

{\it FRB\,190714 and FRB\,200430:} The final positions and uncertainties for these FRBs were determined following the method used for all previous ASKAP/CRAFT FRBs \citep[for detailed descriptions of this method, see][]{Bannister19,Prochaska19b,Macquart20,Day20}. Briefly, the statistical position and uncertainty in R.A. and decl. are derived by fitting a 2D Gaussian to a region containing the FRB in a Stokes $I$ frequency-averaged image. Any errors in the phase-calibration solutions (due to the spatial and temporal differences in the FRB and calibrator observations) are corrected for by comparing the positions of continuum field sources detected in an image made with the 3.1~s of voltage data containing the FRB to positions from reference catalogs, thereby aligning the ASKAP reference frame with the International Celestial Reference Frame (ICRF3). The systematic uncertainties and any offsets are then determined following the method described in \cite{Macquart20}. The statistical FRB position is then corrected for any offset, with the final uncertainty being the quadrature sum of the statistical and systematic uncertainties. 

For FRB\,190714, two quick-look images from the Karl G. Jansky Very Large Array Sky Survey \citep[VLASS,][]{VLASS} were used for comparison with the ASKAP field sources, with offsets and systematic uncertainties in R.A. and decl. determined to be $0\farcs71 \pm 0\farcs32$ and $-1\farcs45 \pm 0\farcs23$, respectively. For comparison with the FRB\,200430 field sources, catalog positions from the Faint Images of the Radio Sky at Twenty centimetres \citep[FIRST,][]{FIRST} survey were used and offsets and uncertainties determined as above. Unusually, FRB\,200340 exhibits a dependence of position in decl. on frequency (yielding an offset in decl. $\approx 7''$ across the frequency band). This indicates a frequency-dependent phase error, potentially due to the ionosphere. Because the FRB and field sources have different spectral indices, their frequency-averaged centroids will likewise differ. In order to account for the bias introduced in correcting the FRB position with the field sources, a coarse spectral index of the FRB was determined by performing a linear fit to the log of the flux densities (extracted from a 56 MHz resolution image cube of the FRB) versus frequency, yielding spectral index $\alpha = -5.46$, and compared to a typical spectral index of the field sources ($\alpha = -0.7$). This was then used to derive the expected deviation in decl. given the offset in the flux-weighted centroid frequencies ($49$~MHz) by evaluating a weighted linear fit of the decl. offsets in the FRB image cube versus frequency at both central frequencies. An offset of $0\farcs93$ was derived, which we also conservatively take to be the typical uncertainty expected due to this bias. Combining this with the offsets and systematic uncertainties derived via the standard field-source comparison method, we obtain a total systematic offset and uncertainty in R.A. and decl. of $-0\farcs03 \pm 0\farcs25$ and $4\farcs12 \pm 1\farcs04$, respectively.

{\it CIGALE:}
As noted in Section~\ref{sec:analysis}, we have reanalyzed
the SED models
of all the previously published hosts from ASKAP/CRAFT
using the same set of model inputs applied
to the new hosts. Because of the sensitivity of
\mstar\ and SFR to assumptions on the SFH and dust,
the new values are quantitatively different.
This is reflected by their large uncertainties, but
we caution the reader that the results are
further subject to systematic errors related to model 
assumptions. In one case (FRB~190608), we also
identified an error in our database that led to
the misreporting of results in \cite{Bhandari20a}.

\subsection{CHIME}

\noindent 
{\it FRB\,180916:} 
\cite{Marcote20} reported the first FRB discovered by the
CHIME experiment \citep{CHIME18} to be localized to 
high-precision, FRB\,180916.J0158+65 (hereafter FRB\,180916
for convenience). It is coincident with the spiral arm
of the previously cataloged galaxy SDSS~J015800.28+654253.0.
As detailed by these authors,
the probability of a chance association is very low,
and our own estimate is $\mpchance = 0.0059$.
We therefore include it in Sample~A.

We also performed SED modeling of the host with {\tt CIGALE}, using archival SDSS optical and mid-infrared WISE photometric data. We compute a stellar mass of $M_\star = (2.15\pm 0.33)\times 10^{9}\,M_{\odot}$, which is approximately a factor of five lower than the estimate by \citet{Marcote20}. We caution that both of these estimates suffer from
large systematic uncertainties due to the substantial corrections
for Galactic extinction.

\subsection{DSA}
\label{sec:dsa}

\noindent 
{\it FRB\,190523:} 

\cite{Ravi19} reported the detection of FRB\,190523 with the
Owens Valley DSA observatory.
The $3'' \times 8''$ 95\%\ error ellipses is nearly coincident
with the centers of two sources labeled S1 and S2 by these authors (at \sone \, and \stwo,\ respectively).
\cite{Ravi19} favor associating S1 with the FRB owing to its
somewhat closer angular offset 
($\theta_{\rm S1} = 3\farcs8$
versus\ $\theta_{\rm S2} \approx 5\farcs1$)
and because its spectroscopic redshift $\mzspec = 0.660$ 
agrees well with the Macquart DM$-z$ relation
\citep{Macquart20}.
Based on the formalism for associations adopted here
(Section~\ref{ssec:associate}), we find
$\mpchance({\rm S1}) = 0.07$ 
and
$\mpchance({\rm S2}) = 0.10$ 

Subsequent to the \cite{Ravi19} publication, we observed
S1 and S2 with the DEIMOS spectrograph on the Keck~II telescope
\citep{Prochaska19c}. The instrument was configured with the
600ZD~grating tilted to cover $\lambda \approx 5000-9500$\AA\ 
and the $1''$ longslit yields a resolution
$R \approx 2500$.
These data were reduced using the \pypeit\ software
package in the same manner as described above.
These data confirm the redshift of S1 reported by
\cite{Ravi19} and yield a spectroscopic redshift for
S2 of $\mzspec = 0.363$ based on H$\alpha$ and [\nii]
nebular emission.

We here revisit the effective prior adopted by \cite{Ravi19}
by adopting the DM$-z$ relation.
These authors report $\mdmfrb = 760.8 \mdmunits$, and the Galactic ISM contribution along this sightline is $\mdmmwism = 37 \, \mdmunits$.  
Assuming a Galactic halo contribution of $\mdmmwhalo = 50 \, \mdmunits$
\citep[][but see \citealt{Keating20}]{Prochaska19a,Platts20}
and a host contribution of $80 \, \mdmunits$ in the host rest-frame,
we have an estimated cosmic dispersion measure 
of $\mdmcosmic({\rm S1}) = 625 \, \mdmunits$
and $\mdmcosmic({\rm S2}) = 616 \, \mdmunits$ for each source.
These are to be compared against the average cosmic dispersion
measure to the redshift of each galaxy,
$\mdmacosmic({\rm S1}) = 607 \, \mdmunits$
and $\mdmacosmic({\rm S2}) = 319 \, \mdmunits$ for each source.
Even allowing for many tens of \dmunits\ uncertainty in the
\dmmwhalo\ and \dmhost\ terms that contribute to the
\dmcosmic\ estimate, the observations favor S1. In any event, neither candidate satisfies
$\mpchance < 0.05$, and we place this association
in Sample~C and associate S1 with the FRB.

\subsection{Realfast}
\label{sec:realfast}

\noindent 
{\it FRB\,190614:} 
\cite{Law20} report the first putative detection of
an FRB from the {\it realfast} collaboration \citep{Law18},
FRB\,20190614D (here referred to as FRB\,190614). They further report the galaxy pair
\acandst\ and \bcandst\ at separations $\approx 1\farcs5$
from the FRB centroid and estimate 
chance probabilities for both galaxies of
$\mpchance(A,B) = 0.07$.
Despite follow-up spectroscopy with the Keck~I telescope,
neither has a secure spectroscopic redshift.
Photometric analysis yields $\mzphot = 0.6$ for each
galaxy with uncertainties of $0.15, 0.2$ for A and B respectively.
These are roughly consistent with the large dispersion
measure reported for FRB\,190614.
In our analysis, we have adopted \bcandst\ as the host,
but we include this system in Sample~D.

\subsection{Others}

\noindent 
{\it FRB\,121102:} 
The host galaxy of FRB\,121102 (also known as ``the Repeater'')
was studied in detail by \cite{Tendulkar17}.
We adopt the majority of their measurements here, but use the updated coordinates for the host galaxy centroid in addition to the estimate of the galaxy's effective half-light radius $R_{\rm eff}$ from \cite{Bassa17}.
The probability of a chance coincidence with the host is 
$\mpchance = 0.002$, and we include this system in Sample~A.

\section{Photometric Data} \label{sec:phot}

The photometric data for the full set of FRB hosts considered in this work are provided in Tables~\ref{tab:photom1}--\ref{tab:photom3}.

\input{tab_photom_sub}

\input{tab_photom_sub2}

\input{tab_photom_sub3}

\end{document}

%% file: tab_sample.tex
\begin{deluxetable*}{cccccccccccccccc}
\tablewidth{0pc}
\tablecaption{Overview of the Main Sample of FRBs and Their Putative Hosts\label{tab:sample}}
\tabletypesize{\footnotesize}
\tablehead{\colhead{FRB}
& \colhead{R.A.$_{\rm FRB}$} & \colhead{Decl.$_{\rm FRB}$} & \colhead{$\sigma_R$}
& \colhead{Repeating}
& \colhead{R.A.$_{\rm host}$} & \colhead{Decl.$_{\rm host}$}
& \colhead{$\theta$}
& \colhead{$\delta x$}
& \colhead{$r_{1/2}$}
& \colhead{$r_i$}
& \colhead{$m$}
& \colhead{Filter} & \colhead{\pchance}
& \colhead{Sample}
\\& (deg) & (deg) & ($''$) & & (deg) & (deg) & ($''$) & ($''$) & ($''$) & ($''$) & (mag)
\\ (1) & (2) & (3) & (4) & (5) & (6) & (7) & (8) & (9) & (10) & (11) & (12) & (13) & (14) & (15)}
\startdata
121102& 82.9946 & $33.1479$& 0.100& y& 82.9946 & $33.1480$& 0.17 & 1.2 & 0.2& 0.44& 23.73 & GMOS\_N\_r& 0.0023& A\\
180916& 29.5031 & $65.7168$& 0.002& y& 29.5012 & $65.7147$& 7.87 & 44.8 & 5.1& 12.95& 16.17 & SDSS\_r& 0.0059& A\\
180924& 326.1052 & $-40.9000$& 0.102& n& 326.1052 & $-40.9002$& 0.71 & 4.7 & 0.6& 1.35& 20.50 & DES\_r& 0.0018& A\\
181112& 327.3485 & $-52.9709$& 1.626& n& 327.3486 & $-52.9709$& 0.28 & 2.7 & 1.2& 3.25& 21.68 & DES\_r& 0.0257& C\\
190102& 322.4157 & $-79.4757$& 0.502& n& 322.4150 & $-79.4757$& 0.45 & 4.1 & 1.0& 2.02& 20.77 & VLT\_FORS2\_I& 0.0050& A\\
190523& 207.0650 & $72.4697$& 2.449& n& 207.0643 & $72.4708$& 3.79 & 2.4 & 0.5& 4.90& 22.01 & Pan-STARRS\_r& 0.0733& C\\
190608& 334.0199 & $-7.8982$& 0.258& n& 334.0204 & $-7.8989$& 3.00 & 20.5 & 1.3& 3.96& 17.55 & SDSS\_r& 0.0016& A\\
190611& 320.7455 & $-79.3976$& 0.671& n& 320.7428 & $-79.3972$& 2.13 & 2.3 & 0.4& 2.27& 22.07 & GMOS\_S\_r& 0.0169& A\\
190614& 65.0755 & $73.7067$& 0.566& n& 65.0738 & $73.7064$& 2.22 & 1.4 & 1.0& 2.99& 23.25 & GMOS\_S\_r& 0.0708& D\\
190711& 329.4195 & $-80.3580$& 0.350& y& 329.4192 & $-80.3581$& 0.49 & 1.3 & 0.5& 1.04& 23.49 & GMOS\_S\_r& 0.0106& A\\
190714& 183.9797 & $-13.0210$& 0.283& n& 183.9796 & $-13.0211$& 0.49 & 4.3 & 1.0& 2.09& 20.69 & Pan-STARRS\_r& 0.0050& A\\
191001& 323.3516 & $-54.7477$& 0.149& n& 323.3519 & $-54.7485$& 2.86 & 13.5 & 1.4& 4.07& 18.34 & DES\_r& 0.0031& A\\
200430& 229.7064 & $12.3769$& 0.546& n& 229.7063 & $12.3766$& 1.04 & 3.0 & 0.6& 1.55& 21.51 & Pan-STARRS\_r& 0.0051& A\\
\hline
\enddata
\tablecomments{Column 1: FRB source. Columns 2 and 3: R.A. and decl. of the FRB
(J2000). Column 4: Approximate FRB localization uncertainty (geometric mean of R.A. and decl. axes). Column 5: FRB classification. Repeating =
yes(y)/no(n). Columns 6 and 7: R.A. and decl. of the associated host galaxy
(J2000). Column 8: Projected angular offset of the FRB to the host-galaxy
center. Column 9: Association radius $\delta x$ \citep{Tunnicliffe14}. Column
10: Angular effective radius of the host measured from a Sérsic model using
GALFIT \citep{galfit} on the $i$-band images (or equivalent). Column 11:
Effective search radius \citep{Bloom02}. Column 12: Measured apparent magnitude
of the host. Column 13: Filter used for the magnitude measurement. Column 14:
Probability of chance coincidence using the \citet{Bloom02} formalism. Column
15: Sample designations following the criteria outlined in
$\S$~\ref{ssec:associate}.}
\end{deluxetable*}

%% file: tab_emlineflux.tex
\begin{deluxetable*}{lccccccc}
\tablewidth{0pc}
\tablecaption{Nebular Emission-line Fluxes\label{tab:lineflux}}
\tabletypesize{\footnotesize}
\tablehead{\colhead{FRB host}
& \colhead{H$\alpha$}
& \colhead{H$\beta$}
& \multicolumn{2}{c}{[O\,\sc{ii}]}
& \multicolumn{2}{c}{[O\,\sc{iii}]}
& \colhead{[N\,{\sc ii}]}
\\& &&  \colhead{$\lambda3726$} & \colhead{$\lambda3729$} & \colhead{$\lambda4959$} & \colhead{$\lambda5007$} & \colhead{$\lambda6584$}
}
\startdata
121102& $2.61\pm 0.04$ & $0.96\pm 0.09$ & --& --& --& $4.38\pm 0.08$ & $<0.12$ \\
180916& $40.3\pm 0.2$ & --& --& --& $5.91\pm 0.62$ & $71.6\pm 0.6$ & $15.2\pm 0.2$ \\
180924& $2.79\pm 0.03$ & $0.72\pm 0.02$ & $0.40\pm 0.02$ & $0.69\pm 0.03$ & --& $0.79\pm 0.02$ & $1.94\pm 0.03$ \\
181112& $0.64\pm 0.30$ & $0.29\pm 0.02$ & --& --& --& $0.54\pm 0.03$ & $0.49\pm 0.30$ \\
190102& $5.66\pm 0.17$ & $1.90\pm 0.17$ & $3.20\pm 0.28$ & $4.21\pm 0.30$ & --& $3.80\pm 0.27$ & $1.69\pm 0.19$ \\
190523& --& $<0.03$ & --& --& --& --& --\\
190608& $27.7\pm 0.4$ & $8.37\pm 0.33$ & $12.1\pm 0.7$ & $19.4\pm 0.8$ & --& $15.0\pm 0.4$ & $18.3\pm 0.4$ \\
190611& $0.49\pm 0.05$ & $0.12\pm 0.03$ & --& --& --& $0.18\pm 0.04$ & $0.12\pm 0.04$ \\
190711& --& $0.26\pm 0.05$ & --& --& --& --& --\\
190714& $3.89\pm 0.03$ & $0.97\pm 0.03$ & --& --& --& $0.31\pm 0.03$ & $1.70\pm 0.03$ \\
191001& $27.4\pm 0.3$ & $5.01\pm 0.30$ & --& --& --& $3.62\pm 0.35$ & $13.9\pm 0.2$ \\
200430& --& --& --& --& --& --& --\\
\hline
\enddata
\tablecomments{Measurements are in units of $10^{-16}$\,erg\,s$^{-1}$\,cm$^{-2}$ and corrected for Galactic dust using the $E(B-V)$ values derived from \citet{Schlafly11}.}
\end{deluxetable*}

%% file: tab_hostprop.tex
\begin{deluxetable*}{cccccccccc}
\tablewidth{0pc}
\tablecaption{Host-galaxy Properties}\label{tab:hostprop}
\tabletypesize{\footnotesize}
\tablehead{\colhead{FRB host} 
& \colhead{$z_{\rm host}$}
& \colhead{$M_{r}$} 
& \colhead{$M_u-M_r$} 
& \colhead{$M_{\star}$} 
& \colhead{SFR} 
& \colhead{Age} 
& \colhead{$Z$} 
& \colhead{Offset} 
& \colhead{$R_{\rm eff}$} 
\\ & & \colhead{(mag)} & \colhead{(mag)} & \colhead{($10^{9}\,M_{\odot}$)} & \colhead{($M_{\odot}$\,yr$^{-1}$)} & \colhead{(Gyr)} & $12+\log$(O/H) & \colhead{(kpc)} & \colhead{(kpc)}
\\ (1) & (2) & (3) & (4) & (5) & (6) & (7) & (8) & (9) & (10)
} 
\startdata 
121102& 0.1927 & $-16.20\pm 0.08$ & $1.49\pm 0.18$ & $0.14\pm 0.07$ & $0.15\pm 0.04$ & $0.26$ & $<8.08$ & $0.6\pm 0.3$ & $0.7\pm 0.1$ \\
180916& 0.0337 & $-19.46\pm 0.05$ & $1.53\pm 0.06$ & $2.15\pm 0.33$ & $0.06\pm 0.02$ & $0.15$ & -- & $5.5\pm 0.1$ & $3.6\pm 0.4$ \\
180924& 0.3212 & $-20.81\pm 0.05$ & $1.78\pm 0.15$ & $13.2\pm 5.1$ & $0.88\pm 0.26$ & $0.38$ & $8.93^{+0.02}_{-0.02}$ & $3.4\pm 0.5$ & $2.7\pm 0.1$ \\
181112& 0.4755 & $-20.40\pm 0.07$ & $1.12\pm 0.15$ & $3.98\pm 2.02$ & $0.37\pm 0.11$ & $0.57$ & $8.86^{+0.10}_{-0.13}$ & $1.7\pm 19.2$ & $7.2\pm 1.7$ \\
190102& 0.2912 & $-19.85\pm 0.06$ & $1.40\pm 0.12$ & $3.39\pm 1.02$ & $0.86\pm 0.26$ & $0.06$ & $8.70^{+0.07}_{-0.08}$ & $2.0\pm 2.2$ & $4.4\pm 0.5$ \\
190523& 0.6600 & $-22.06\pm 0.12$ & $1.92\pm 0.19$ & $61.2\pm 40.1$ & $<0.09^{a}$ & $0.69$ & -- & $27\pm 23$ & $3.3\pm 0.2$ \\
190608& 0.1178 & $-21.22\pm 0.05$ & $1.40\pm 0.09$ & $11.6\pm 2.8$ & $0.69\pm 0.21$ & $0.38$ & $8.85^{+0.02}_{-0.02}$ & $6.6\pm 0.6$ & $2.8\pm 0.2$ \\
190611& 0.3778 & -- & -- & $\sim 0.8$ & $0.27\pm 0.08$ & -- & $8.71^{+0.17}_{-0.28}$ & $11\pm 4$ & $2.1\pm 0.1$ \\
190711& 0.5220 & $-19.01\pm 0.08$ & $0.95\pm 0.16$ & $0.81\pm 0.29$ & $0.42\pm 0.12^{a}$ & $0.61$ & -- & $3.2\pm 2.1$ & $2.9\pm 0.2$ \\
190714& 0.2365 & $-19.92\pm 0.05$ & $1.19\pm 0.17$ & $14.9\pm 7.1$ & $0.65\pm 0.20$ & $1.59$ & $9.03^{+0.04}_{-0.04}$ & $1.9\pm 1.1$ & $3.9\pm 0.1$ \\
191001& 0.2340 & $-22.13\pm 0.05$ & $1.67\pm 0.19$ & $46.4\pm 18.8$ & $8.06\pm 2.42$ & $0.64$ & $8.94^{+0.05}_{-0.05}$ & $11\pm 1$ & $5.5\pm 0.1$ \\
200430& 0.1600 & $-18.05\pm 0.05$ & $1.78\pm 0.31$ & $1.30\pm 0.60$ & $\sim 0.2^{b}$ & $0.69$ & -- & $3.0\pm 2.4$ & $1.6\pm 0.5$ \\
\hline 
\enddata 
\tablecomments{Column 1: FRB source. Column 2: Host redshift. Spectroscopic redshifts are reported to four significant digits (typical uncertainty), and photometric redshifts to two significant digits. Column 3: Absolute $r$-band magnitude. Column 4: Rest-frame $M_u-M_r$ colors. Column 5: Stellar mass from SED modeling. Column 6: Star-formation rate derived from the line luminosity of \ha, assuming a \citet{Chabrier03} IMF. Column 7: Estimate of the mass-weighted stellar population age from SED modeling. Column 8: Oxygen abundance derived using the O3N2 calibration from \cite{Hirschauer18}. Column 9: Projected physical offset of the FRB from the galaxy center. Column 10: Effective radius of the host galaxy measured in the $i$ band (or equivalent).}
\tablenotetext{a}{The SFR is derived from \hb\ assuming a nominal scaling with \ha\ (i.e., no internal host extinction).}
\tablenotetext{b}{The SFR is derived from the best-fit photometric SED model.}
\end{deluxetable*} 

%% file: tab_photom_sub.tex
\begin{deluxetable*}{ccccccccccccc}
\rotate
\tablewidth{0pc}
\tablecaption{PHOTOMETRY.\label{tab:photom1}}
\tabletypesize{\footnotesize}
\tablehead{\colhead{Filter}  
& \colhead{HG121102} 
& \colhead{HG180916} 
& \colhead{HG180924} 
& \colhead{HG181112} 
& \colhead{HG190102} 
& \colhead{HG190523} 
& \colhead{HG190608} 
& \colhead{HG190611} 
& \colhead{HG190711} 
& \colhead{HG190714} 
& \colhead{HG191001} 
& \colhead{HG200430} 
} 
\startdata 
\cutinhead{DES}
$g$ &  &  & 21.56 & 22.64 &  &  &  &  &  &  & 19.12 & \\
$\sigma (g)$ &  &  & 0.03 & 0.09 &  &  &  &  &  &  & 0.00 & \\
$r$ &  &  & 20.50 & 21.68 &  &  &  &  &  &  & 18.34 & \\
$\sigma (r)$ &  &  & 0.02 & 0.05 &  &  &  &  &  &  & 0.00 & \\
$i$ &  &  & 20.11 & 21.46 &  &  &  &  &  &  & 17.91 & \\
$\sigma (i)$ &  &  & 0.02 & 0.06 &  &  &  &  &  &  & 0.00 & \\
$z$ &  &  & 19.83 & 21.42 &  &  &  &  &  &  & 17.74 & \\
$\sigma (z)$ &  &  & 0.02 & 0.11 &  &  &  &  &  &  & 0.00 & \\
$Y$ &  &  & 19.79 & 21.05 &  &  &  &  &  &  & 17.63 & \\
$\sigma (Y)$ &  &  & 0.06 & 0.17 &  &  &  &  &  &  & 0.01 & \\
\cutinhead{Pan-STARRS}
$g$ &  &  &  &  &  & 22.92 &  &  &  & 21.20 &  & 22.16\\
$\sigma (g)$ &  &  &  &  &  & 0.17 &  &  &  & 0.04 &  & 0.08\\
$r$ &  &  &  &  &  & 22.01 &  &  &  & 20.69 &  & 21.51\\
$\sigma (r)$ &  &  &  &  &  & 0.10 &  &  &  & 0.03 &  & 0.06\\
$i$ &  &  &  &  &  & 21.14 &  &  &  & 20.38 &  & 21.16\\
$\sigma (i)$ &  &  &  &  &  & 0.06 &  &  &  & 0.02 &  & 0.04\\
$z$ &  &  &  &  &  & 20.79 &  &  &  & 20.05 &  & 20.91\\
$\sigma (z)$ &  &  &  &  &  & 0.06 &  &  &  & 0.03 &  & 0.07\\
$y$ &  &  &  &  &  & 20.59 &  &  &  & 20.04 &  & 20.67\\
$\sigma (y)$ &  &  &  &  &  & 0.10 &  &  &  & 0.05 &  & 0.18\\
\cutinhead{SDSS}
$u$ &  & 20.31 &  &  &  &  & 18.99 &  &  &  &  & \\
$\sigma (u)$ &  & 1.78 &  &  &  &  & 0.09 &  &  &  &  & \\
$g$ &  & 17.08 &  &  &  &  & 18.02 &  &  &  &  & \\
$\sigma (g)$ &  & 0.08 &  &  &  &  & 0.02 &  &  &  &  & \\
$r$ &  & 16.17 &  &  &  &  & 17.55 &  &  &  &  & \\
$\sigma (r)$ &  & 0.03 &  &  &  &  & 0.01 &  &  &  &  & \\
$i$ &  & 15.93 &  &  &  &  & 17.22 &  &  &  &  & \\
$\sigma (i)$ &  & 0.02 &  &  &  &  & 0.02 &  &  &  &  & \\
$z$ &  & 15.85 &  &  &  &  & 17.09 &  &  &  &  & \\
$\sigma (z)$ &  & 0.06 &  &  &  &  & 0.05 &  &  &  &  & \\
\hline 
\enddata 
\tablecomments{All photometry has been corrected for Galactic extinction.} 
\end{deluxetable*}

%% file: tab_photom_sub2.tex
\begin{deluxetable*}{ccccccccccccc}
\rotate
\tablewidth{0pc}
\tablecaption{PHOTOMETRY.\label{tab:photom2}}
\tabletypesize{\footnotesize}
\tablehead{\colhead{Filter}  
& \colhead{HG121102} 
& \colhead{HG180916} 
& \colhead{HG180924} 
& \colhead{HG181112} 
& \colhead{HG190102} 
& \colhead{HG190523} 
& \colhead{HG190608} 
& \colhead{HG190611} 
& \colhead{HG190711} 
& \colhead{HG190714} 
& \colhead{HG191001} 
& \colhead{HG200430} 
} 
\startdata 
\cutinhead{WISE}
$W1$ &  & 14.37 & 16.84 &  &  &  & 14.37 &  &  &  &  & \\
$\sigma (W1)$ &  & 0.03 & 0.10 &  &  &  & 0.03 &  &  &  &  & \\
$W2$ &  & 14.41 & 16.06 &  &  &  & 13.83 &  &  &  &  & \\
$\sigma (W2)$ &  & 0.05 & 0.18 &  &  &  & 0.04 &  &  &  &  & \\
$W3$ &  & 10.56 & 11.69 &  &  &  & 10.76 &  &  &  &  & \\
$\sigma (W3)$ &  & 0.08 & -999.00 &  &  &  & 0.12 &  &  &  &  & \\
$W4$ &  & 9.08 & 8.50 &  &  &  & 8.65 &  &  &  &  & \\
$\sigma (W4)$ &  & 0.52 & -999.00 &  &  &  & 0.41 &  &  &  &  & \\
\cutinhead{GMOS\_N}
$g$ & 23.33 &  &  &  &  &  &  &  &  &  &  & \\
$\sigma (g)$ & 0.12 &  &  &  &  &  &  &  &  &  &  & \\
$r$ & 23.73 &  &  &  &  &  &  &  &  &  &  & \\
$\sigma (r)$ & 0.14 &  &  &  &  &  &  &  &  &  &  & \\
$i$ & 23.54 &  &  &  &  &  &  &  &  &  &  & \\
$\sigma (i)$ & 0.09 &  &  &  &  &  &  &  &  &  &  & \\
$z$ & 23.49 &  &  &  &  &  &  &  &  &  &  & \\
$\sigma (z)$ & 0.13 &  &  &  &  &  &  &  &  &  &  & \\
\cutinhead{GMOS\_S}
$g$ &  &  &  &  &  &  &  &  & 23.47 &  &  & \\
$\sigma (g)$ &  &  &  &  &  &  &  &  & 0.20 &  &  & \\
$r$ &  &  &  &  &  &  &  & 22.07 & 23.49 &  &  & \\
$\sigma (r)$ &  &  &  &  &  &  &  & 0.15 & 0.15 &  &  & \\
$i$ &  &  &  &  &  &  &  & 22.34 & 22.95 &  &  & \\
$\sigma (i)$ &  &  &  &  &  &  &  & 0.15 & 0.15 &  &  & \\
\cutinhead{VLT}
$u$ &  &  &  &  & 22.77 &  &  &  &  &  &  & \\
$\sigma (u)$ &  &  &  &  & 0.20 &  &  &  &  &  &  & \\
$g$ &  &  & 21.32 & 22.50 & 21.87 &  &  &  &  & 20.47 & 18.89 & \\
$\sigma (g)$ &  &  & 0.04 & 0.04 & 0.10 &  &  &  &  & 0.10 & 0.10 & \\
$I$ &  &  & 20.07 & 21.48 & 20.77 &  &  &  &  & 19.50 & 17.84 & \\
$\sigma (I)$ &  &  & 0.02 & 0.04 & 0.05 &  &  &  &  & 0.10 & 0.10 & \\
$z$ &  &  &  &  & 20.54 &  &  &  &  &  &  & \\
$\sigma (z)$ &  &  &  &  & 0.20 &  &  &  &  &  &  & \\
\hline 
\enddata 
\tablecomments{All photometry has been corrected for Galactic extinction.} 
\end{deluxetable*}

%% file: tab_photom_sub3.tex
\begin{deluxetable*}{ccccccccccccc}
\rotate
\tablewidth{0pc}
\tablecaption{PHOTOMETRY.\label{tab:photom3}}
\tabletypesize{\footnotesize}
\tablehead{\colhead{Filter}  
& \colhead{HG121102} 
& \colhead{HG180916} 
& \colhead{HG180924} 
& \colhead{HG181112} 
& \colhead{HG190102} 
& \colhead{HG190523} 
& \colhead{HG190608} 
& \colhead{HG190611} 
& \colhead{HG190711} 
& \colhead{HG190714} 
& \colhead{HG191001} 
& \colhead{HG200430} 
} 
\startdata 
\cutinhead{Spitzer}
$3.6$ & 23.79 &  &  &  &  &  &  &  &  &  &  & \\
$\sigma (3.6)$ & 0.20 &  &  &  &  &  &  &  &  &  &  & \\
$4.5$ & 24.72 &  &  &  &  &  &  &  &  &  &  & \\
$\sigma (4.5)$ & 999.00 &  &  &  &  &  &  &  &  &  &  & \\
\cutinhead{VISTA}
$Y$ &  &  &  &  &  &  &  &  &  & 18.01 &  & \\
$\sigma (Y)$ &  &  &  &  &  &  &  &  &  & 0.12 &  & \\
$J$ &  &  &  &  &  &  &  &  &  & 17.56 &  & \\
$\sigma (J)$ &  &  &  &  &  &  &  &  &  & 0.09 &  & \\
$H$ &  &  &  &  &  &  &  &  &  & 17.06 &  & \\
$\sigma (H)$ &  &  &  &  &  &  &  &  &  & 0.10 &  & \\
$Ks$ &  &  &  &  &  &  &  &  &  & 16.47 &  & \\
$\sigma (Ks)$ &  &  &  &  &  &  &  &  &  & 0.20 &  & \\
\cutinhead{WFC3}
F110W & 23.08 &  &  &  &  &  &  &  &  &  &  & \\
$\sigma$ (F110W) & 0.01 &  &  &  &  &  &  &  &  &  &  & \\
F160W & 22.96 &  &  &  & 20.45 &  &  &  & 22.73 & 18.88 &  & \\
$\sigma$ (F160W) & 0.03 &  &  &  & 0.01 &  &  &  & 0.01 & 0.00 &  & \\
\hline 
\enddata 
\tablecomments{All photometry has been corrected for Galactic extinction.} 
\end{deluxetable*}